\documentclass[12pt]{article}

\usepackage[latin1]{inputenc}
\usepackage[T1]{fontenc}
\usepackage{color}
\usepackage{eurosym}
\usepackage[top=4cm,bottom=4cm,left=3cm,right=3cm]{geometry}
\usepackage{fancyhdr}
\usepackage{float}
\usepackage{amssymb,amsmath, amsthm}
\usepackage{url,graphicx}

\usepackage{booktabs}
\usepackage{stmaryrd} 
\usepackage{mathenv}
\usepackage{dsfont}
\usepackage{amssymb}
\usepackage{latexsym}
\usepackage{amsmath}
\usepackage{color}
\usepackage{comment}
\usepackage{amsthm}% for

\newcommand{\reels}{\mathbb{R}}
\newcommand{\naturels}{\mathbb{N}}

\newcommand{\esp}{\mathbb{E}}
\newcommand{\proba}{\mathbb{P}}
\newcommand{\var}{\operatorname{Var}}

\newcommand{\Tau}{\mathrm{T}}

\theoremstyle{plain}

\newtheorem{clt}{Theorem}
\newtheorem{cltreg}[clt]{Theorem}
\newtheorem{thQMLE}[clt]{Theorem}
\newtheorem{thEQMLE}[clt]{Theorem}
\newtheorem{thEQMLEpowers}[clt]{Theorem}
\newtheorem{cltexampleuncertainty}[clt]{Theorem}
\newtheorem{exMA1}[clt]{Theorem}

\theoremstyle{remark}
\newtheorem{continuousass}{Remark}

\newtheorem{notIIDexamples}[continuousass]{Remark}
\newtheorem{conditionC1rk}[continuousass]{Remark}

\newtheorem{cltblocksize}[continuousass]{Remark}

\newtheorem{AVrk}[continuousass]{Remark}
\newtheorem{biasGtrk}[continuousass]{Remark}
\newtheorem{biasGtrknonreg}[continuousass]{Remark}
\newtheorem{h-}[continuousass]{Remark}

\newtheorem{rkuncertaintycvrate}[continuousass]{Remark}

\newtheorem{rkC1}[continuousass]{Remark}

\newtheorem{rkE1}[continuousass]{Remark}
\newtheorem{rkparammodel}[continuousass]{Remark}

\newtheorem{rkhighfrequencycov}[continuousass]{Remark}

\theoremstyle{remark}
\newtheorem{notIID}{Example}

\newtheorem{notIID2}[notIID]{Example}

\newtheorem{Fn}[notIID]{Example}
\newtheorem{Uijn}[notIID]{Example}
\newtheorem{estim1}[notIID]{Example}
\newtheorem{estim2}[notIID]{Example}

\theoremstyle{definition}
\newtheorem*{conditionC1}{Condition (C)}

\newtheorem*{conditionT}{Condition (T)}
\newtheorem*{conditionE}{Condition (E)}
\newtheorem*{conditionE*}{Condition (E*)}
\newtheorem*{conditionP1}{Condition (P1)}
\newtheorem*{conditionP2}{Condition (P2)}

\newtheorem*{conditionE3'}{Condition (E3')}
\newtheorem*{conditionE5'}{Condition (E5')}
\newtheorem*{conditionE6'}{Condition (E6')}
\newtheorem*{stableconvergence}{Definition}

\begin{document}

 \title{Local Parametric Estimation in High Frequency Data\footnote{We are indebted to Simon Clinet, Takaki Hayashi, Dacheng Xiu, participants of the seminars in Berlin and Tokyo and conferences in Osaka, Toyama, the SoFie annual meeting in Hong Kong, the PIMS meeting in Edmonton for valuable comments, which helped in improving the quality of the paper.}}
 \author{Yoann Potiron\footnote{Faculty of Business and Commerce, Keio University. 2-15-45 Mita, Minato-ku, Tokyo, 108-8345, Japan. Phone:  +81-3-5418-6571. E-mail: potiron@fbc.keio.ac.jp website: http://www.fbc.keio.ac.jp/\char`\~ potiron} and Per Mykland\footnote{Department of Statistics, The University of Chicago. 
5734 S. University Avenue,
Chicago IL, 60637. Phone:  + 1 773 702 8044/8333. Fax: + 1 773 702 9810. E-mail: mykland@pascal.uchicago.edu}}
\date{This version: \today}

\maketitle

\begin{abstract}
In this paper, we give a general time-varying parameter model, where the multidimensional parameter possibly includes jumps. The quantity of interest is defined as the integrated value over time of the parameter process $\Theta = T^{-1} \int_0^T \theta_t^* dt$. We provide a local parametric estimator (LPE) of $\Theta$ and conditions under which we can show the central limit theorem. Roughly speaking those conditions correspond to some uniform limit theory in the parametric version of the problem. The framework is restricted to the specific convergence rate $n^{1/2}$. Several examples of LPE are studied: estimation of volatility, powers of volatility, volatility when incorporating trading information and time-varying MA(1).
\end{abstract}

\textbf{Keywords}: integrated volatility; market microstructure noise; powers of volatility; quasi maximum likelihood estimator

% high frequency data; local parametric estimator

\section{Introduction}

Modeling dynamics is essential in various fields, including finance, economics, physics, environmental 
engineering, geology and sociology. Time-varying parametric models can deal with a specific problem in dynamics, namely, the temporal evolution of systems. The extensive literature on time-varying parameter models and local parametric methods include and are not limited to Fan and Gijbels (1996), Hastie and Tibshirani (1993) or Fan and Zhang (1999)  when regression and generalized regression models are involved, locally stationary processes following the work of Dahlhaus (1997, 2000), Dahlhaus and Rao (2006), or any other time-varying parameter models, e.g. Stock and Watson (1998) and Kim and Nelson (2006). 

\smallskip
In this paper, we propose to specify local parametric methods in the particular context of high-frequency statistics for a broad class of problems. Local methods have been used extensively in the high-frequency data literature, see e.g. Mykland and Zhang (2009, 2011), Kristensen (2010), Rei\ss { }(2011) or Jacod and Rosenbaum (2013) among many others. If we define $T$ as the horizon time, the (random) target quantity in this monograph is defined as the integrated parameter 
\begin{eqnarray}
\label{integral} \Theta := \frac{1}{T} \int_0^T \theta_s^* ds,
\end{eqnarray} 
which can be equal to the volatility, the covariation between several assets, the variance of the microstructure noise, the friction parameter of the model with uncertainty zones (see Example \ref{p2uncertaintyzones} for more details), 
the time-varying parameters of the MA(1) model, etc. To estimate the integrated parameter, we estimate the \emph{local parameter} on each block by using the parametric estimator on the observations within the block and take a weighted sum of the local parameter estimates, where each weight is equal to the corresponding block length. We call the obtained estimator the \emph{local parametric estimator} (LPE).

\smallskip
In Section \ref{secCLT}, we investigate conditions under which we can establish the related central limit theorem with convergence rate $n^{1/2}$, where $n$ is the (possibly expected) number of observations. The framework is such that the local block length vanishes asymptotically. Basically, we aim to provide the statistician with a transparent and as simple as possible device to tackle the time-varying parameter problem based on central limit theory in the parametric version of the problem. The original key probabilistic step of the proof, which formally allows for switching from random to deterministic parameter, is the use of regular conditional distribution theory (see, e.g., Breiman (1992)). The price to pay is some kind of uniformity in the parametric limit theory results, and to show that some deviation between the parametric and the time-varying parameter model vanishes asymptotically.

\smallskip
In Section \ref{examples}, the technology is used on five distinct examples to derive the related central limit theorems. As far as the authors know, all those results are new. Depending on the considered example, the LPE is useful for one or several of the following  reasons: 
{\renewcommand\labelitemi{}
\begin{itemize}
  \item \emph{robustness}: the LPE is robust to time-varying parameters (such as the noise variance, $\eta$ from the model with uncertainty zones, the parameters of the MA(1) process) which are usually assumed constant. This is the case of all our examples, except for Example \ref{exyingying}.
  \item \emph{efficiency}: the LPE turns out to be more efficient than the global estimator or existing concurrent approaches. This is the case of Example \ref{exyingying}. In addition, the LPE is conjectured to be efficient in all our examples except for Example \ref{p2uncertaintyzones}.
  \item \emph{definition of new estimators}: It can be the case that the estimator does not work globally but that the LPE provides a good candidate as in Example \ref{expowers} and Example \ref{exyingying}.
\end{itemize}
}

\smallskip
We describe the five examples in what follows. To estimate integrated volatility under noisy observations, Xiu (2010) studied the quasi-maximum likelihood-estimator (QMLE) originally examined in A\"{i}t-Sahalia et al. (2005), showed the corresponding asymptotic theory when the variance of the noise is fixed and obtained a convergence rate $n^{1/4}$, which is optimal (see Gloter and Jacod (2001)). More recently, A\"{i}t-Sahalia and Xiu (2016) establish that it is robust to shrinking noise satisfying $O_p(1/n^{3/4})$ and Da and Xiu (2017) obtains central limit theorem with rate ranging from $n^{1/4}$ to $n^{1/2}$ depending on the magnitude of the noise. When assuming that it is $O_p(1/\sqrt{n})$, we show that the LPE of the QMLE is optimal (with rate $n^{1/2}$) and furthermore robust to time-varying noise variance.

\smallskip
Another important problem, which goes back to Barndorff-Nielsen and Shephard (2002), is the estimation of higher powers of volatility. To do that, we define a LPE where the local estimates are powers of the QMLE of volatility. Under the assumption on small noise $O_p(1/\sqrt{n})$, we show that this estimator is optimal and robust to time-varying noise variance. This is an example where the global approach does not work as the QMLE is only consistent when estimating volatility.

\smallskip
A more recent problem is the estimation of volatility when incorporating trading information. To do that, Li et al. (2016) assume that the noise is a parametric function
of trading information with a remaining noise component of order $O_p(1/\sqrt{n})$. Their strategy consists in first estimating the parametric part of the noise, and then take the sum of square pre-estimated efficient returns. They also advocate for the use of the QMLE after price pre-estimation although they do not provide the associated limit theory. We show that the latter approach, when considering the LPE of QMLE, is optimal and provides a better asymptotic variance (AVAR) than the former technique. In addition, a modification of the local estimator as in Example 2 allows us to estimate higher powers of volatility.

\smallskip
A concurrent ultra high frequency approach to model the observed price was given in Robert and Rosenbaum (2011, 2012), who introduced the semiparametric model with uncertainty zones where $\eta$ is the 1-dimensional friction parameter, observation times are endogenous and observed prices lie on a tick grid. As most likely correlated with the volatility, it is natural to consider $\eta_t$ as a time-varying parameter. We provide a formal model extension and establish the according limit theory of the LPE of the estimator considered in their work. In addition, our empirical illustration seems to indicate that $\eta_t$ is indeed time-varying.

\smallskip
In the last example, we consider an application in time series and introduce a time-varying MA(1) model with null mean. The time series is observed in high frequency on $[0,T]$ and $\theta_t^*$ corresponds to the two-dimensional parameter of the MA(1) process. We show that the LPE of the MLE is optimal and document that it outperforms the global MLE and other concurrent approaches in finite sample. 

\smallskip
The remaining of this paper is organized as follows. The LPM is introduced in the following section. Conditions for the central limit theory are stated in Section 3. We give the examples in Section 4. We investigate the finite sample performance of the local QMLE of volatility and compares it to the global approach in Section 5. We conclude in Section 6. Consistency in a simple model, proofs, additional numerical simulations on MA(1) model and an empirical illustration on the model with uncertainty zones are gathered in Appendix.

\section{The Locally Parametric Model (LPM)} \label{LPM}
\subsection{Data-generating mechanism}
We assume that we observe the  $d$-dimensional vectors $Z_{0,n},$ $\cdots , Z_{N_n,n}$, where $N_n$ can be random, the observation times satisfy $\tau_{0,n} := 0 < \tau_{1,n} < \cdots < \tau_{N_n,n} \leq T$. The observations and the observation times are both related to the latent parameter $\theta_t^*$. 

\smallskip
As an example, the observations can satisfy $Z_{\tau_{i,n},n} = X_{\tau_{i,n}} + \epsilon_{i,n}$, where $X_t = \sigma_t dW_t$ stands for the efficient price, $W_t$ is a standard Brownian motion, $\epsilon_{i,n}$ corresponds to the market microstructure noise (which will be restricted to be of order $\epsilon_{i,n} = O_p (1/\sqrt{n})$ due to the limitation of the technology developed in Section \ref{secCLT}), is independent and identically distributed (IID) and independent from $X_t$, and the latent parameter is equal to the volatility, i.e. $\theta_t^* = \sigma_t^2$.

We assume that the parameter process $\theta_t^*$ takes values in $K$, a (not necessarily compact) subset of $\reels^p$.  
We do not assume any independence between $\theta_t^*$ and the other quantities driving the observations, such as the Brownian motion of the efficient price process. In particular, there can be leverage 
effect (see e.g. Wang and Mykland (2014), A\"{i}t-Sahalia et al. (2017)). Also, the arrival times $\tau_{i,n}$ and the parameter $\theta_t^*$ can be correlated, i.e. there is (some kind of) endogeneity in sampling times.

\subsection{Asymptotics}
There are commonly two choices of asymptotics in the literature: the \emph{high-frequency} asymptotics, 
which makes the number of observations explode on $[0,T]$, and the \emph{low-frequency} asymptotics, which takes $T$ to infinity. We choose the former one. Investigating the low-frequency implementation case 
is beyond the scope of this paper\footnote{If we set down the asymptotic theory in the same way as in p.3 of Dahlhaus (1997), we conjecture that the results of this paper would stay true.}.

\subsection{Estimation}
The approach taken here is frequent in high-frequency data. We define the block size (i.e. the number of observations in a block) as $h_n$, and the number of blocks as $B_n := \ulcorner N_n h_n^{-1} \urcorner$. For $i=1, \cdots, B_n$ we define the parameter average on the $ith$ block as 
\begin{eqnarray}
\Theta_{i,n} := \frac{\int_{\Tau_{i-1,n}}^{\Tau_{i,n}} \theta_s^* ds}{\Delta \Tau_{i,n}},
\end{eqnarray}
where $\Tau_{i,n} := \min ( \tau_{i h_n}, T )$ and its corresponding parametric estimator as $\widehat{\Theta}_{i,n}$. Then, we take the weighted 
sum of $\widehat{\Theta}_{i,n}$ and obtain an estimator of the integrated spot process
\begin{eqnarray}
\label{newestimator} \widehat{\Theta}_n := \frac{1}{T} \sum_{i=1}^{B_n} \widehat{\Theta}_{i,n} \Delta \Tau_{i,n},
\end{eqnarray}
where $\Delta \Tau_{i,n} = \Tau_{i,n} - \Tau_{i-1,n}$. We call (\ref{newestimator}) the \emph{local parametric estimator} (LPE). We assume that 
\begin{eqnarray}
h_n/n & \rightarrow & 0
\end{eqnarray}
so that when observations are regular the block size $\Delta \Tau_{i,n} := Th_n/n$ vanishes asymptotically. In view of Condition (T) and Remark \ref{rkE1}, we have similarly that $\esp [\Delta \Tau_{i,n}] = O(h_n/n)$ also goes to 0 when observations are not regular. 

\section{The central limit theorem} \label{secCLT}
We present in this section the general technology of our paper.\footnote{Note that the local approach in this paper is related to the large-T-based-approach and problem of Giraitis et al. (2014).} It is mainly based on Theorem 2-2 in Jacod (1997), or similarly Theorem IX.7.3 and Theorem IX.7.28 in Jacod and Shiryaev (2003) or Theorem 2.2.15 in Jacod and Protter (2011), along with regular conditional distribution techniques (see, e.g., Section 4.3 (pp. 77-80) in Breiman (1992)). More specifically, we provide sufficient conditions to the aforementioned theorem in the particular context of this paper. Those conditions are based on the limit theory in the parametric version of the problem, which we assume pre-obtained by the statistician.

\smallskip
The following methods are specified\footnote{It is possible to specify the problem with a general rate of convergence, but all the considered examples from this paper are with convergence rate $n^{1/2}$.} to the rate of convergence $n^{\frac{1}{2}}$. Formally, we aim to find the limit distribution of
\begin{eqnarray}
\label{goal} n^{\frac{1}{2}} T^{-1} \sum_{i=1}^{B_n} \big( \widehat{\Theta}_{i,n} - \Theta_{i,n} \big) \Delta \Tau_{i,n}.
\end{eqnarray}
Specifically, we want to show that (\ref{goal}) converges \emph{stably}\footnote{One can look at definitions of 
stable convergence in R\'{e}nyi (1963), Aldous and Eagleson (1978), Chapter 3 (p. 56) of Hall and Heyde (1980), 
Rootz\'{e}n (1980), Section 2 (pp. 169-170) of Jacod and Protter (1998), Definition VIII.5.28 in Jacod and Shiryaev (2003) or Definition 1 in Podolskij and Vetter (2010).} to a limit distribution. We first give the definition of stable convergence.
\begin{stableconvergence}
(Stable convergence) A sequence of random variables $Z_{n}$ is said to converge $\mathcal{J}_T$-stably to $Z$, which is defined on an extension $(\Omega', \mathcal{F}', P')$ of $(\Omega, \mathcal{F}, P)$, if for any $E \in \mathcal{J}_T$ and for any  continuous bounded function $f$ we have
$$\esp \big[ f(Z_n) \mathbf{1}_{E} \big] \rightarrow \esp' \big[ f(Z) \mathbf{1}_{E} \big].$$
\end{stableconvergence}

\subsection{Regular observation case}
We consider first the simple case when observations are regular, i.e. $\tau_{i,n} = iT/n$ and $N_n = n$. We assume that $\mathcal{J}_{t}$ is a (continuous-time) filtration on $(\Omega, \mathcal{F}, P)$ such that 
$\theta_t^*$ is adapted to it. In the following of this paper, when using the conditional expectation $\esp_{\tau} [ Z ]$\footnote{The related assumption is that $\tau$ is a $\mathcal{J}_t$-stopping time.}, we will refer to the conditional expectation of $Z$ knowing 
$\mathcal{J}_{\tau}$. We define the discrete-time version of the filtration as $\mathcal{I}_{i,n} = \mathcal{J}_{\tau_{i,n}}$. Finally, if we denote the returns of the observations as \begin{eqnarray}
 R_{i,n} = Z_{\tau_{i,n},n} - Z_{\tau_{i-1,n},n},
\end{eqnarray}
we assume that the returns can be expressed as
\begin{eqnarray} \label{timevarparammodel}
 \label{recret} R_{i,n} = F_n \big( \{P_{s,n}\}_{0 \leq s \leq \tau_{i-1,n}}, U_{i,n}, \{ \theta_s^* \}_{\tau_{i-1,n} \leq s \leq \tau_{i,n}}\big),
\end{eqnarray}
where $F_n(x,y,z)$ is a $\reels^{d}$-dimensional non-random function\footnote{We assume that  $F_n(x,y,z)$ is jointly measurable, and that $P_{t,n}$ is taking values on a Borel space. Additionally, we assume that for any $( P_{s,n}, U_{i,n}, \theta_s^*)$, we have 
$\esp \mid F_n (P_{s,n}, U_n, \theta_s^*) \mid < \infty$}, the random innovation $U_{i,n}$ are IID (although with distribution which can depend on $n$) adapted to $\mathcal{I}_{i,n}$ and independent of the past information $\mathcal{I}_{i-1,n}$, $P_{t,n}$ is a (possibly multidimensional) process adapted to $\mathcal{J}_{t}$ which stands for the past that matters in the model. We further assume that $P_{t,n}$ is independent from  $\theta_t^*$.

\smallskip
The key example stands as follows. We assume that the observations are following the additive model $Z_{\tau_{i,n},n} = X_{\tau_{i,n}} + \epsilon_{i,n}$, where $X_t = \sigma_t dW_t$ is the efficient price and $\epsilon_{i,n}$ the (shrinking) IID noise independent from $X_t$, and that the parameter is $\theta_t^* = \sigma_t^2$. In that case $U_{i,n} = (\{ W_s \}_{\tau_{i-1,n} \leq s \leq \tau_{i,n}} - W_{\tau_{i-1,n}}, \epsilon_{i,n})$, and $P_{s,n} = \epsilon_{i,n}$ if $\tau_{i,n} \leq s < \tau_{i+1,n}$. The function\footnote{The advised reader will have noticed that $F_n$ is not a function in the ordinary sense. We still abusively refer to it as a "function".} $F_n$ takes on the form
\begin{eqnarray}
\label{Fnkeyexample}
F_n = \int_{\tau_{i-1,n}}^{\tau_{i,n}} \sigma_s dW_s + \epsilon_{i,n} - \epsilon_{i-1,n}.
\end{eqnarray}
Crucial to the expression (\ref{Fnkeyexample}) is that the dependence in the past is only through the past noise $\epsilon_{i-1,n}$, i.e. we do not need to know the whole past of $P_{t,n}$, but rather only the current value. This will be very useful in what follows.

\smallskip
We provide now the outline of the method. Our goal is to investigate the limit distribution of (\ref{goal}) using prior limit result on the parametric version of the problem. A common approach in high frequency statistics proofs consists in decomposing $\big( \widehat{\Theta}_{i,n} - \Theta_{i,n} \big)$ into 
\begin{eqnarray}
\label{decomposeCLT} 
\big(\widehat{\Theta}_{i,n}  - \hat{\tilde{\Theta}}_{i,n} \big) 
+ \big( \hat{\tilde{\Theta}}_{i,n} - \theta_{\Tau_{i-1,n}}^{*} \big) + \big( \theta_{\Tau_{i-1,n}}^{*} - \Theta_{i,n} \big),
\end{eqnarray} 
where $\hat{\tilde{\Theta}}_{i,n}$ stands for the estimator when we hold the parameter constant on each block. Then, one can usually deal with the first term and the third term (most likely using Burkholder-Davis-Gundy and Markov type of inequalities) and eventually show that they vanish asymptotically. The main work lies in establishing the central limit theory of the second term in (\ref{decomposeCLT}). A typical proof consists in using locally parametric results along with some Riemann sum argument. But this can be cumbersome as the parameter on each block, although constant, is random. Instead, we propose to look at the further decomposition of $\big( \hat{\tilde{\Theta}}_{i,n} - \theta_{\tau_{i-1,n}}^{*} \big)$ into
\begin{eqnarray}
 \big( \hat{\tilde{\Theta}}_{i,n} -  \hat{\tilde{\Theta}}_{i,n}^{\mathbf{P}} \big) + \big( \hat{\tilde{\Theta}}_{i,n}^{\mathbf{P}} - \theta_{\tau_{i-1,n}}^{*} \big) \text{, where } \label{thetadecompo} \\ \label{thetaP}
 \hat{\tilde{\Theta}}_{i,n}^{\mathbf{P}} :=  \hat{\tilde{\Theta}}_{i,n} \mid  \{P_{s,n}\}_{0 \leq s \leq \tau_{i-1,n}} = \mathbf{P},
\end{eqnarray}
and $\mathbf{P}$ is a fixed non-random past. In the case of (\ref{Fnkeyexample}), we can choose $\mathbf{P} = 0$. From this new decomposition, it is expected as relatively accessible to show that the first term goes to 0, so that the central limit theory will be investigated on the second term of the decomposition. By conditioning by one particular past in (\ref{thetaP}), we got rid of some randomness, although the parameter is still random. Using conditional regular distribution results in our proofs, we actually show that we can also take the parameter non-random. The price to pay for such method is to show some kind of uniformity in the parameter value when showing the limit results, and that the first term in (\ref{thetadecompo}) vanishes asymptotically. 

\smallskip
We introduce some definition. For $i=1, \cdots, B_n$ we define the returns on the $i$th block $R_{i,n}^j := 
R_{(i-1) h_n + j,n}$ for $j=1, \cdots, h_n$, and similarly $U_{i,n}^j$, $\tau_{i,n}^j$, $W_{i,n}^j$ and $\epsilon_{i,n}^j$. We assume that
\begin{eqnarray} \label{Thetahat}
 \widehat{\Theta}_{i,n} := \hat{\theta}_{h_n,n} (R_{i,n}^{1} ; \cdots ; 
 R_{i,n}^{h_n}),
\end{eqnarray}
where $\hat{\theta}_{h_n,n}$ is a function on $\reels^{dh_n}$. The approximated returns and the approximated estimates are defined as
\begin{eqnarray}
 \label{appreturnsdef} \widetilde{R}_{i,n}^j & := &  F_n \big( \{P_{s,n}\}_{0 \leq s \leq \tau_{i,n}^{j-1}}, U_{i,n}^j,  \theta_{\Tau_{i-1,n}}^* \big),\\
 \hat{\tilde{\Theta}}_{i,n} & := & \hat{\theta}_{h_n,n} (\widetilde{R}_{i,n}^{1} ; \cdots ; 
 \widetilde{R}_{i,n}^{h_n}). 
\end{eqnarray}
Basically, those two expressions can be seen as the pendant of respectively (\ref{timevarparammodel}) and (\ref{Thetahat}) when we hold the parameter constant equal to its block initial value $\theta_{\Tau_{i-1,n}}^*$. In the case of the key example (\ref{Fnkeyexample}), we obtain that the approximated returns are of the form
\begin{eqnarray}
 \widetilde{R}_{i,n}^j & = & \sigma_{\Tau_{i-1,n}}(W_{i,n}^j - W_{i,n}^{j-1}) + (\epsilon_{i,n}^j - \epsilon_{i,n}^{j-1}).
\end{eqnarray}
We also introduce the conditional parametric version as
\begin{eqnarray}
\label{conditionalparamdef} \widetilde{R}_{i,n}^{j,\mathbf{P}} & := & \esp \big[ \widetilde{R}_{i,n}^{j} \mid \{U_{i,n}^k \}_{k \leq j}, \{P_{s,n}\}_{0 \leq s \leq \Tau_{i-1,n}} = \mathbf{P} \big],\\
 \hat{\tilde{\Theta}}_{i,n}^{\mathbf{P}} & := & \hat{\theta}_{h_n,n} (\widetilde{R}_{i,n}^{1,\mathbf{P}} ; \cdots ; 
 \widetilde{R}_{i,n}^{h_n,\mathbf{P}}).
\end{eqnarray}
Here, we fix the past equal to $\mathbf{P}$ in (\ref{conditionalparamdef}), which removes some randomness compared with (\ref{appreturnsdef}). In the key example, we can (arbitrarily) choose $\mathbf{P}=0$, and this past will only "affect" the first conditional parametric version of the return on the block equal to \begin{eqnarray}
 \widetilde{R}_{i,n}^{1,\mathbf{P}} & = & \sigma_{\Tau_{i-1,n}}(W_{i,n}^1 - W_{i,n}^{0}) + \epsilon_{i,n}^1,
\end{eqnarray}
whereas for $j= 2, \cdots, h_n$, we have $ \widetilde{R}_{i,n}^{j,\mathbf{P}} =  \widetilde{R}_{i,n}^{j}$. This key example is an instance where the model is 1-Markovian in the sense that the past only affects the value of the first return on the block. This is quite mild assumption, and we will see that more sophisticated models, such as the model with uncertainty zones, naturally exhibit longer past time-dependence.
Moreover, we introduce a parametric version of the returns and the estimators when the parameter is equal to $\theta$ and the past fixed to $\mathbf{P}$. Accordingly, the randomness is further reduced in the following expressions. This will be useful in Condition (E).
\begin{eqnarray}
R_{i,n}^{j,\mathbf{P},\theta} & := & \esp \big[ \widetilde{R}_{i,n}^{j} \mid \{U_{i,n}^k \}_{k \leq j}, \theta_{\Tau_{i-1,n}}^{*} = \theta, \{P_{s,n}\}_{0 \leq s \leq \Tau_{i-1,n}} = \mathbf{P} \big],\\
\hat{\tilde{\Theta}}_{i,n}^{\mathbf{P},\theta} & := & \hat{\theta}_{h_n,n} (R_{i,n}^{1,\mathbf{P},\theta} ; \cdots ; 
 R_{i,n}^{h_n,\mathbf{P},\theta}). \label{ThetatildehatPtheta}
\end{eqnarray}

\smallskip
We provide now the assumptions on $\theta_t^*$. The first assumption considers the continuous It\^{o}-semimartingale case.
\begin{conditionP1}
The parameter $\theta_t^*$ is of the form 
\begin{eqnarray}
\label{paramito} d \theta_t^* := a_t^{\theta} dt +  \sigma_t^{\theta} dW_t^{\theta},
\end{eqnarray}
where $a_t^{\theta}$ is adapted locally bounded (of dimension $p$) and $\sigma_t^{\theta}$ is a non-negative continuous It\^{o}-process adapted locally bounded (of dimension $p \times p$),
and $W_t^{\theta}$ is a standard {$p$}-dimensional Brownian motion.
\end{conditionP1}
We introduce a norm for $u \in \reels^p$ as $\mid u \mid = \sqrt{(u^{(1)})^2 + \cdots + (u^{(p)})^2}$. The following assumption allows for a more general process than semi-martingales.  Nonetheless, this assumption is quite restrictive, in particular since $h_n$ does not show up on the right hand-side of (\ref{P2eq}). In practice this is useful when considering a smooth parameter which cannot be expressed as a "pure drift".
\begin{conditionP2}
$\theta_t^*$ satisfies uniformly in $i=1, \cdots, B_n$ that 
\begin{eqnarray}
\label{P2eq}\esp_{\Tau_{i-1,n}} \big[ \underset{\Tau_{i-1,n} \leq s \leq \Tau_{i,n}}{\sup} 
 \mid \theta_s^* - \theta_{\Tau_{i-1,n}}^* \mid^2  
 \big] = o_p (n^{-1}).
\end{eqnarray}
\end{conditionP2}
As the uniformity of limit results on the whole space $K$ might be impossible to obtain, we allow to work on the compact subspace $K_M$, which grows to $K$ as $M$ increases. Accordingly, we assume that $\theta_t^*$ is locally bounded on a compact set $K_M$ in the sense that there exists $\tau_m \overset{\proba}{\rightarrow} T$ such that for any $m$, there exists $M_m > 0$ which satisfies $\theta_t^* \in K_{M_m}$ for any $t \in [0,\tau_m]$.

\smallskip
We provide in what follows sufficient conditions to the bias condition (3.10), the increment condition (3.11) and the Lindeberg condition (3.13) in Theorem 3-2 from Jacod (1997). (Almost) equivalently, Theorem IX.7.3 and Theorem IX.7.28 in Jacod and Shiryaev (2003) or Theorem 2.2.15 in Jacod and Protter (2011) could have been used. Those conditions are based on the parametric version of the problem.
\begin{conditionE}
For any (non random) parameter $\theta \in K$, we assume that there exists a (non random) covariance matrix $V_{\theta}$  positive definite such that for any $M > 0$, we have $V_{\theta}$ is bounded for any  $\theta \in K_M$ and uniformly in $\theta \in K_M$ and in $i=1, \cdots, B_n$ we have
\begin{eqnarray}
\label{E6110}  \esp \big[ \big( \hat{\tilde{\Theta}}_{i,n}^{\mathbf{P},\theta} - \theta \big) \big]  & = & o( n^{-\frac{1}{2}})\\
  \var \big[ h_n^{\frac{1}{2}} \big( \hat{\tilde{\Theta}}_{i,n}^{\mathbf{P},\theta} - \theta \big) \big]  & = & V_{\theta} T  + o (1) \label{E3210}\\
\label{E3310} \esp \big[ h_n \mid \hat{\tilde{\Theta}}_{i,n}^{\mathbf{P},\theta} - \theta  \mid^2 \mathbf{1} \{ h_n n^{-\frac{1}{2}} \mid \hat{\tilde{\Theta}}_{i,n}^{\mathbf{P},\theta} - \theta  \mid > \epsilon \} \big]  & = & o (1) \text{ , } \forall \epsilon > 0.
\end{eqnarray}
\end{conditionE}

\smallskip
We let $B_n(t)$ be the number of blocks before $t$, and $\mathcal{M}_b$ the set of all bounded martingales. We now provide the central limit theorem.
\begin{cltreg} (Central limit theorem with regular observation times)
\label{cltreg}
 We assume Condition (E). Moreover, we assume Condition (P1) and that the block size $h_n$ is such that
 \begin{eqnarray}
  \label{E21} n^{- \frac{1}{2}} h_n & = & o(1),
 \end{eqnarray}
or Condition (P2). Let $M_t$ be a $p$-dimensional square-integrable continuous martingale. Furthermore, we assume that for all $t \in [0,T]$ we have
 \begin{eqnarray}
  \label{Gt} n^{-\frac{1}{2}} h_n \sum_{i=1}^{B_n(t)} \esp_{\Tau_{i-1,n}} \Big[\big( \hat{\tilde{\Theta}}_{i,n}^{\mathbf{P}} - \theta_{\tau_{i-1,n}}^* \big) \big( M_{\Tau_{i,n}} - M_{\Tau_{i-1,n}} \big)^T \Big] & \overset{\proba}{\rightarrow} & 0,\\
  \label{Nt} n^{-\frac{1}{2}} h_n \sum_{i=1}^{B_n(t)} \esp_{\Tau_{i-1,n}} \Big[\big( \hat{\tilde{\Theta}}_{i,n}^{\mathbf{P}} - \theta_{\tau_{i-1,n}}^* \big) \big( N_{\Tau_{i,n}} - N_{\Tau_{i-1,n}} \big) \Big] & \overset{\proba}{\rightarrow} & 0,
 \end{eqnarray}
 for all $N \in \mathcal{M}_b(M^{\perp})$, where $\mathcal{M}_b(M^{\perp})$ is the class of all elements of $\mathcal{M}_b$ which are orthogonal to $M$ (i.e. to all components of $M$). Finally, we assume that 
 \begin{eqnarray} \label{condE+}
  n^{-\frac{1}{2}} h_n \sum_{i=1}^{B_n} \big( \widehat{\Theta}_{i,n} - \hat{\tilde{\Theta}}_{i,n}^{\mathbf{P}} \big)   & \overset{\proba}{\rightarrow} & 0.
 \end{eqnarray}
 Then, stably in law as $n \rightarrow \infty$, we have
 \begin{eqnarray}
  \label{cltconclusion} n^{\frac{1}{2}} \big( \widehat{\Theta}_n - \Theta \big) \rightarrow \widetilde{Z}, 
  \end{eqnarray}
  where $\langle \widetilde{Z} ,\widetilde{Z} \rangle_t = T^{-1} \int_0^t V_{\theta_s^*} ds$, and $\langle \widetilde{Z} ,M \rangle_t = 0$. In particular, we have
\begin{eqnarray}
 \label{maincltreg} n^{\frac{1}{2}} \big( \widehat{\Theta}_n - \Theta \big) \rightarrow \Big(T^{-1} \int_0^T V_{\theta_s^*} ds \Big)^{\frac{1}{2}} \mathcal{N} (0,1).
\end{eqnarray}
\end{cltreg}

\begin{rkparammodel}
(Parametric model) Note that in the case where the time-varying parameter model is equal to the parametric model with parameter equal to $\theta^*$, the asymptotic variance (AVAR) of $\widehat{\Theta}_n$ is equal to the variance of the parametric model, i.e.
$$n^{\frac{1}{2}} \big( \widehat{\Theta}_n - \Theta \big) \rightarrow V_{\theta^*}^{\frac{1}{2}} \text{ } \mathcal{N} (0,1).$$
\end{rkparammodel}

\begin{AVrk} (Estimating the asymptotic variance) If the statistician does not have a (parametric) variance estimator at hand and that her parametric estimator can be written as in Mykland and Zhang (2017), one can use 
the techniques of the cited paper to obtain a variance estimate. Investigating if such techniques would work in our setting is beyond the scope of this paper. If she has a variance estimator $\hat{v}_{h_n,n}$, then for any $i=1, \cdots, B_n$ she can estimate 
the $i$th block variance $\widehat{V}_{i,n}$ as $\widehat{V}_{i,n} := \hat{v}_{h_n,n} (R_{i,n}^{1} ; \cdots ; R_{i,n}^{h_n})$, and the asymptotic variance as the weighted sum
\begin{eqnarray}
\label{AV} \widehat{V}_n = T^{-1} \sum_{i=1}^{B_n} \widehat{V}_{i,n} \Delta \Tau_{i,n}.
\end{eqnarray} 
\end{AVrk}
This estimator will be consistent under mild uniformity assumptions.

\begin{biasGtrk} (Non-zero asymptotic bias)
If we further assume that in place of condition (\ref{Gt}) there is a non-zero continuous process $G_t$ such that 
\begin{eqnarray}
  n^{-\frac{1}{2}} h_n \sum_{i=1}^{B_n(t)} \esp_{\Tau_{i-1,n}} \Big[\big( \hat{\tilde{\Theta}}_{i,n}^{\mathbf{P}} - \theta_{\tau_{i-1,n}}^* \big) \big( M_{\Tau_{i,n}} - M_{\Tau_{i-1,n}} \big)^T \Big] & \overset{\proba}{\rightarrow} & G_t,
  \end{eqnarray}
  then (\ref{cltconclusion}) still holds, where $\langle \widetilde{Z} ,\widetilde{Z} \rangle_t = T^{-1} \int_0^t V_{\theta_s^*} ds$ and $\langle \widetilde{Z} ,M \rangle_t = G_t$, but (\ref{maincltreg}) no longer holds. 
\end{biasGtrk}
\subsection{Non regular observation case}
We consider now the case when observations can be random (even endogenous). We define the increment of time as $\Delta \tau_{i,n} := \tau_{i,n} - \tau_{i-1,n}$ and make the first natural assumption.
\begin{conditionT}
The observation times are such that 
  \begin{eqnarray}
  \esp \big[N_n \big] &= &O(n),\\
  \label{obs1} \underset{1 \leq i \leq N_n}{\sup} \text{ } \esp_{\tau_{i-1,n}} \big[ ( \Delta \tau_{i,n} )^3 \big] 
  & = & O_p (n^{-3}).
  \end{eqnarray}
\end{conditionT}
\begin{rkE1} \label{rkE1} (block length) As an obvious consequence of (\ref{obs1}), we have that the block length satisfies $\esp \big[ \Delta \Tau_{i,n} \big] = O (h_n n^{-1})$. 
\end{rkE1}
The observation times are related to $\theta_t^*$, as are the returns. We assume that $(R_{i,n}, \Delta \tau_{i,n})$ satisfies (\ref{timevarparammodel}), and that all the definitions (\ref{Thetahat}) - (\ref{ThetatildehatPtheta}) follow. Finally, we define $\Delta \widetilde{\Tau}_{i,n}^{\mathbf{P}} = \widetilde{\tau}_{h_n i,n}^{\mathbf{P}} - \widetilde{\tau}_{(h_n-1)i,n}^{\mathbf{P}}$ and $\Delta \Tau_{i,n}^{\mathbf{P},\theta} = \tau_{h_n i,n}^{\mathbf{P},\theta} - \tau_{(h_n-1)i,n}^{\mathbf{P},\theta}$. We adapt Condition (E) in this case.
\begin{conditionE*}
For any (non random) parameter $\theta \in K$, we assume that there exists a (non random) covariance matrix $V_{\theta} > 0$ such that for any $M > 0$, we have $V_{\theta}$ is bounded for any  $\theta \in K_M$ and uniformly in $\theta \in K_M$ and in $i=1, \cdots, B_n$ we have 
\begin{eqnarray}
\label{E611}  \esp \big[ \big( \hat{\tilde{\Theta}}_{i,n}^{\mathbf{P},\theta} - \theta \big) \Delta \Tau_{i,n}^{\mathbf{P},\theta}\big]  & = & o( h_n n^{-\frac{3}{2}}),\\
  \var \big[ h_n^{\frac{1}{2}} \big( \hat{\tilde{\Theta}}_{i,n}^{\mathbf{P},\theta} - \theta \big) \Delta \Tau_{i,n}^{\mathbf{P},\theta} \big]  & = & V_{\theta} \esp \big[ \Delta \Tau_{i,n}^{\mathbf{P},\theta} \big] T h_n n^{-1} \label{E321}\\ \nonumber & & + o (h_n^2 n^{-2}), \\
\label{E331} \esp \big[ n^2 h_n^{-1} (A_{i,n}^{\mathbf{P},\theta} )^2 \mathbf{1}_{\{ n^{\frac{1}{2}} A_{i,n}^{\mathbf{P},\theta} > \epsilon \}} \big] & = & o (1) \text{ , } \forall \epsilon > 0,
\end{eqnarray}
where $A_{i,n}^{\mathbf{P},\theta} = \mid \hat{\tilde{\Theta}}_{i,n}^{\mathbf{P},\theta} - \theta  \mid \Delta \Tau_{i,n}^{\mathbf{P},\theta}$.
\end{conditionE*}
We also adapt the central limit theorem.
\begin{clt} (Central limit theorem with non regular observation times)
\label{clt}
 We assume Condition (T) and Condition (E*). Moreover, we assume Condition (P1) and (\ref{E21}),
or Condition (P2). Let $M_t$ be a p-dimensional square-integrable continuous martingale. Furthermore, we assume that for all $t \in [0,T]$ we have
 \begin{eqnarray}
  \label{condnotreg1} \frac{n^{\frac{1}{2}}}{T} \sum_{i=1}^{B_n(t)} \esp_{\Tau_{i-1,n}} \Big[\big( \hat{\tilde{\Theta}}_{i,n}^{\mathbf{P}} - \theta_{\tau_{i-1,n}}^* \big) \Delta \widetilde{\Tau}_{i,n}^{\mathbf{P}} \big( M_{\Tau_{i,n}} - M_{\Tau_{i-1,n}} \big)^T \Big] & \overset{\proba}{\rightarrow} & 0,\\ \label{condnotreg2}
  n^{\frac{1}{2}} \sum_{i=1}^{B_n(t)} \esp_{\Tau_{i-1,n}} \Big[\big( \hat{\tilde{\Theta}}_{i,n}^{\mathbf{P}} - \theta_{\tau_{i-1,n}}^* \big) \Delta \widetilde{\Tau}_{i,n}^{\mathbf{P}} \big( N_{\Tau_{i,n}} - N_{\Tau_{i-1,n}} \big) \Big] & \overset{\proba}{\rightarrow} & 0,
 \end{eqnarray}
 for all $N \in \mathcal{M}_b(M^{\perp})$. Finally, we assume that 
 \begin{eqnarray}
  \label{nonregeq} n^{\frac{1}{2}} \sum_{i=1}^{B_n} \big( \widehat{\Theta}_{i,n} \Delta \Tau_{i,n} - \hat{\tilde{\Theta}}_{i,n}^{\mathbf{P}} \Delta \widetilde{\Tau}_{i,n}^{\mathbf{P}} \big)  & \overset{\proba}{\rightarrow} & 0,\\
  \label{nonregeq2} n^{\frac{1}{2}} \sum_{i=1}^{B_n} \esp_{\Tau_{i-1,n}} \Big[ \big| \Delta \widetilde{\Tau}_{i,n}^{\mathbf{P}} - \Delta \Tau_{i,n} \big| \Big] & \overset{\proba}{\rightarrow} & 0,
 \end{eqnarray}
 uniformly in $i=1,\cdots,B_n$. Then, stably in law as $n \rightarrow \infty$, we have
 \begin{eqnarray}
 \label{cltnonreg0} n^{\frac{1}{2}} \big( \widehat{\Theta}_n - \Theta \big) \rightarrow \widetilde{Z}, 
  \end{eqnarray}
  where $\langle \widetilde{Z} ,\widetilde{Z} \rangle_t = T^{-1} \int_0^t V_{\theta_s^*} ds$, and $\langle \widetilde{Z} ,M \rangle_t = 0$. In particular, we have
\begin{eqnarray}
 \label{mainclt} n^{\frac{1}{2}} \big( \widehat{\Theta}_n - \Theta \big) \rightarrow \Big(T^{-1} \int_0^T V_{\theta_s^*} ds \Big)^{\frac{1}{2}} \mathcal{N} (0,1).
\end{eqnarray}
\end{clt}

\begin{biasGtrknonreg} (Non-zero asymptotic bias) More generally, if we assume that there is a non-zero continuous process $G_t$ such that for all $t \in [0,T]$ we have
 \begin{eqnarray}
  \frac{n^{\frac{1}{2}}}{T} \sum_{i=1}^{B_n(t)} \esp_{\Tau_{i-1,n}} \Big[\big( \hat{\tilde{\Theta}}_{i,n}^{\mathbf{P}} - \theta_{\tau_{i-1,n}}^* \big) \Delta \widetilde{\Tau}_{i,n}^{\mathbf{P}} \big( M_{\Tau_{i,n}} - M_{\Tau_{i-1,n}} \big)^T \Big] & \overset{\proba}{\rightarrow} & G_t,
  \end{eqnarray}
  instead of (\ref{condnotreg1}), then (\ref{cltnonreg0}) still holds, where $\langle \widetilde{Z} ,\widetilde{Z} \rangle_t = T^{-1} \int_0^t V_{\theta_s^*} ds$, and $\langle \widetilde{Z} ,M \rangle_t = G_t$, but (\ref{mainclt}) no longer holds.
  \end{biasGtrknonreg}

\subsection{Bias correction}
As the parametric estimator must satisfy the bias condition (\ref{E611}), it is useful to consider in some instances a bias-corrected (BC) version of it which provides the estimate on the ith block $\widehat{\Theta}_{i,n}^{(BC)}$. The BC LPE is then constructed as
$$\widehat{\Theta}_{n}^{(BC)} = \frac{1}{T} \sum_{i=1}^{B_n} \widehat{\Theta}_{i,n}^{(BC)} \Delta \Tau_{i,n}.$$

\section{Examples}
\label{examples}
This section provides some applications of the theory introduced in Section \ref{secCLT}. The central limit theorems provided in this section are all new. We choose four examples with regular observations in which it is sufficient to show the conditions of Theorem \ref{cltreg}. We further consider the model with uncertainty zones where there is endogeneity in observation times implying that we have to verify the more general conditions of Theorem \ref{clt}.

\subsection{Estimation of volatility with the QMLE}
\label{noise}
\subsubsection{Central limit theorem}
We assume that the noise has the form $$\epsilon_{i,n} := n^{-\alpha} v_{\tau_{i,n}}^{\frac{1}{2}} \gamma_{\tau_{i,n}},$$ 
where $\alpha \geq 1/2$, the noise variance $v_t$ is time-varying, and $\gamma_t$ are IID with null-mean and unity variance. In other words we have $\epsilon_{i,n} = O_p ( 1/\sqrt{n})$. The parameter process is defined as the two-dimensional volatility and noise variance process $\theta_t^* = (\sigma_t^2, v_t)$ and thus $\Theta = \big(T^{-1} \int_0^T \sigma_t^2 dt, T^{-1} \int_0^T v_t dt \big) $. Correspondingly we work locally with the QMLE considered in Xiu (2010, p. 236) and we introduce the notation for the corresponding LPE $\widehat{\Theta}_n = (\widehat{\sigma}_n^2, \widehat{v}_n)$. 

\smallskip
We also consider the bias-corrected version of the QMLE $\widehat{\Theta}_n^{(BC)}$, where the procedure to construct the unbiased estimator is given in Section \ref{unbiasQMLE}. In numerical simulations under a realistic framework, this bias is not observed even with small values of $n$ (see Section 6 in Xiu (2010) and Section 5 in Clinet and Potiron (2018b)), and thus it is safe to use $\widehat{\Theta}_n = (\widehat{\sigma}_n^2, \widehat{v}_n)$ in practice. 

\smallskip
The assumption of $\alpha \geq 1/2$ is quite restrictive in view of the related literature on the QMLE. Unfortunately in the case $\alpha < 1/2$, the techniques of this paper do not apply. Xiu (2010) showed the CLT of the QMLE when $v_t$ is non time-varying and $\alpha=0$. In the same setting Clinet and Potiron (2018b) showed that the asymptotic variance can be smaller when using the LPE with $B_n = B$ fixed and documented that in finite sample the LPE was advantageous over the global QMLE. A\"{i}t-Sahalia and Xiu (2016) actually establish that the MLE is robust to noise of the form $O_p(1/n^{3/4})$. Da and Xiu (2017) show the central limit theory with rate of convergence ranging from $n^{1/2}$ to $n^{1/4}$ depending on the magnitude of the noise. 

\smallskip
However the techniques allow us to investigate how the LPE behaves in a different asymptotics, i.e. when the noise variance is $O_p ( 1/\sqrt{n})$ and $B_n$ tends to $+ \infty$. Moreover, we allow for heteroskedasticity in noise variance. Finally, in the case where the noise variance goes to 0 at the same speed as the variance of the returns, i.e. $\alpha= 1/2$, we can also retrieve the integrated variance noise. In accordance with the setting of this paper, the convergence rate of both the volatility and the noise is $n^{1/2}$.

\smallskip
To verify the conditions for the CLT, we use heavily the asymptotic results of the QMLE (see Theorem 6 in Xiu (2010)) and the MLE in the low-frequency asymptotics (see Proposition 1 on p. 369 in A\"{i}t-Sahalia et al. (2005)). The result is formally embedded in the following theorem.
\begin{thQMLE} (QMLE)
\label{thQMLE}
 We define $\mathcal{F}_t^X$ the filtration generated by $X_t$. 
 
 (i) We assume that $\alpha > \frac{1}{2}$. Then, $\mathcal{F}_T^X$-stably in law as $n \rightarrow \infty$,
\begin{eqnarray}
\label{thQMLE1}
 n^{\frac{1}{2}} \Big( \widehat{\sigma}_n^{2} - T^{-1} \int_0^T \sigma_s^2 ds \Big) \rightarrow \Big(6 T^{-1} \int_0^T \sigma_s^4 ds \Big)^{\frac{1}{2}} \mathcal{N} (0,1).
\end{eqnarray}

(ii) When $\alpha = \frac{1}{2}$, we have $\mathcal{F}_T^X$-stable convergence in law of $n^{\frac{1}{2}} \big( \widehat{\Theta}_n^{(BC)} - \Theta \big)$ to a mixed normal random variable with zero mean and asymptotic variance given by 
\begin{eqnarray}
\label{thQMLE2}
  T^{-1} \left(\begin{matrix} A & - \int_0^T \big(\sigma_s^4 + 2 \sigma_s^2 v_s + 4 \sigma_s^3 \sqrt{4v_s + \sigma_s^2} \big) ds\\ 
\bullet & \frac{1}{2} \int_0^T \big(2 v_s + \sigma_s^2 \big) \big(\sigma_s^2 + 2 v_s + \sigma_s \sqrt{4v_s + \sigma_s^2} \big) ds
\end{matrix}\right),
\end{eqnarray}
where $A = \int_0^T \big(2 \sigma_s^4 + 4 \sigma_s^3 \sqrt{4v_s + \sigma_s^2} \big) ds$.
\end{thQMLE}

\begin{rkhighfrequencycov} (Estimation of high-frequency covariance with the QMLE) To estimate integrated covariance under noisy observations and asynchronous observations, A\"{i}t-Sahalia et al. (2010) introduced a QMLE based on a synchronization of observation times. It is clear that their Generalized Synchronization Method can be expressed as a LPM. In view of the close connection between their proposed estimator (2) on p. 1506 and the QMLE studied in Section \ref{noise}, the conditions of our work can be verified and thus Theorem 2 (p. 1506) of the authors can be adapted with the LPE in a framework similar to Section \ref{noise}, i.e. when the noise variance is $O(n^{-1/2})$ and time-varying.
\end{rkhighfrequencycov}

\subsubsection{Algorithm to construct the unbiased estimator}
\label{unbiasQMLE}
We describe here the algorithm to obtain $\widehat{\Theta}_n^{(BC)}$. Note that the bias-correction is only required when $\alpha=1/2$.
\begin{enumerate}
    \item We compute the local QMLEs.
    \item From Theorem 6 (p. 241) in Xiu (2010), we compute the corresponding $W_1$ and $W_2$.
    \item We change some entries of the matrices to ensure unbiased estimates when using formula (21) and (22) in the aforementioned theorem.
    \item  We compute the unbiased local QMLE using the formula (21) and (22) with the corrected matrices.
    \item The bias-corrected LPE $\widehat{\Theta}_n^{(BC)}$ is taken as the mean of local bias-corrected estimates.
\end{enumerate}

\subsection{Estimation of powers of volatility}
\label{expowers}
Here the parameter is $\theta_t^* = g(\sigma_t^2)$ with $g$ not being the identity function. We are concerned with the estimation of powers of volatility $\Theta = T^{-1} \int_0^T g(\sigma_t^2) dt$ under microstructure noise with variance $O ( 1/\sqrt{n})$ in the same setting as in Section \ref{noise}. 

\smallskip
The problem was introduced in Barndorff-Nielsen and Shephard (2002). They showed that the case $g(x) = x^2$ is related to the asymptotic
variance of the realized volatility. One can also consult Barndorff-Nielsen et al. (2006), Mykland and Zhang (2012, Proposition 2.17, p. 138) and Renault et al. (2017) for related developments. All those studies assume no microstructure noise. 

\smallskip
When there is microstructure noise, Jacod et al. (2010) use the pre-averaging method. In the special case of quarticity, one can also look at Mancino and Sanfelici (2012) and Andersen et al. (2014). In the case of tricicity, see Altmeyer and Bibinger (2015).

\smallskip
Under no microstructure noise, block estimation (Mykland and Zhang (2009, Section 4.1, p. 1421-1426)) has the ability to make
the mentioned estimators approximately or fully efficient. The path followed to do that is to first estimate locally the volatility $\widehat{\sigma}_{i,n}^2$ and then take a Riemann sum of $g(\widehat{\sigma}_{i,n}^2)$. See also Jacod and Rosenbaum (2013) for an extended version of the method in some ways.

\smallskip
In the same spirit when allowing for microstructure noise, we propose to use locally the estimation $g(\widehat{\sigma}_{i,n}^2)$, where $\widehat{\sigma}_{i,n}^2$ is the QMLE estimate of the volatility on the $i$th block. As pointed out in Jacod and Rosenbaum (2013), even if we use locally the bias-corrected estimator $\big(\widehat{\sigma}_{i,n}^{(BC)}\big)^2$, we will pay a price for the fact that we use the function $g$ in front. In particular, an asymptotic bias quite challenging to correct for will appear in the asymptotic limit theory, as seen in Theorem 3.1 in the cited paper. To get rid of most parts of this bias, we follow the idea at the beginning of Section 3.2 of the cited work and choose $h_n$ such that
\begin{eqnarray}
\label{powerass} n^{-1/2} h_n^{3/2} \rightarrow \infty.
\end{eqnarray}
Note that this is not incompatible with the other condition (\ref{E21}), i.e. $n^{-1/2} h_n \rightarrow 0$, that will be assumed in what follows. With (\ref{powerass}), the part of the bias that doesn't vanish grows to the extent that it explodes asymptotically. This leads us to consider the following two bias-corrected estimators:
\begin{eqnarray}
\label{1608}\widehat{\Theta}_n^{(BC,1)} & = & B_n^{-1} \sum_{i=1}^{B_n} \big( g(\widehat{\sigma}_{i,n}^2) - \frac{3}{h_n} \widehat{\sigma}_{i,n}^4 g''(\widehat{\sigma}_{i,n}^2) \big).\\
\widehat{\Theta}_n^{(BC,2)} & = & B_n^{-1} \sum_{i=1}^{B_n} \Bigg( g\Big(\big(\widehat{\sigma}_{i,n}^{(BC)}\big)^2\Big) \nonumber \\ \label{1609} & &- \frac{( \widehat{\sigma}_{i,n}^{(BC)})^4 + 2 ( \widehat{\sigma}_{i,n}^{(BC)})^3\sqrt{4\widehat{v}_{i,n}^{(BC)} + (\widehat{\sigma}_{i,n}^{(BC)})^2 }}{h_n}g''\Big(\big(\widehat{\sigma}_{i,n}^{(BC)}\big)^2\Big)  \Bigg).
\end{eqnarray}
The theorem is given in what follows. The proof  uses a local delta method and then follows the proof of Theorem \ref{thQMLE}.

\begin{thQMLE} (powers of volatility)
\label{thQMLEpowers} Let $g$ a nonnegative function such that 
\begin{eqnarray}
\mid g^{(j)}(x) \mid \leq K(1 + \mid x \mid^{p-j}), \text{ } j=0,1,2,3,
\end{eqnarray}
for some constants $K>0,p\geq3$.\\
(i) We assume that $\alpha > \frac{1}{2}$. Then, $\mathcal{F}_T^X$-stably in law as $n \rightarrow \infty$,
\begin{eqnarray}
 n^{\frac{1}{2}} \big( \widehat{\Theta}_n^{(BC,1)} - \Theta \big) \rightarrow \Big(6 T^{-1} \int_0^T \big( g' (\sigma_s^2) \big)^2 \sigma_s^4 ds \Big)^{\frac{1}{2}} \mathcal{N} (0,1).
\end{eqnarray}
(ii) When $\alpha = \frac{1}{2}$, we have $\mathcal{F}_T^X$-stably in law that
\begin{eqnarray*}
\label{thQMLEpowers1}
 n^{\frac{1}{2}} \big( \widehat{\Theta}_n^{(BC,2)} - \Theta \big) \rightarrow 
 \Big( T^{-1} \int_0^T \big( g' (\sigma_s^2) \big)^2 \big(2 \sigma_s^4 + 4 \sigma_s^3 \sqrt{4v_s + \sigma_s^2} \big) ds \Big)^{\frac{1}{2}} \mathcal{N} (0,1).
\end{eqnarray*}
\end{thQMLE}

To reflect on the powerfulness of the local approach, the reader can note that the global QMLE is estimating the wrong quantity when $g$ is different from the identity function, except when the volatility is constant. To see why this is the case, we consider the estimation of quarticity (i.e. with $g(x)=x^2$) and we note that a global QMLE would estimate $g(\int_0^T \sigma_t^2 dt)$, which is except when volatility is constant different from $\int_0^T \sigma_t^4 dt$. The extensive empirical work in Andersen et al. (2014) also indicates that the two quantities are very different in practice.  

\subsection{Estimation of volatility and higher powers of volatility incorporating trading information}
\label{exyingying}
To incorporate all the information  available in high frequency data (e.g. in addition to transaction prices, we also observe the trading volume, the type of trade i.e. buyer or seller initiated, more generally bid/ask information from the limit order book), Li et al. (2016) consider the model where the noise is partially observed through a parametric function $$Z_{\tau_{i,n},n} = X_{\tau_{i,n}} + \epsilon_{i,n} = X_{\tau_{i,n}} +  h(I_{i,n}, \nu) + \tilde{\epsilon}_{i,n},$$ where $I_{i,n}$ is the vector of information at time $\tau_{i,n}$ and $\tilde{\epsilon}_{i,n}$ is the noisy part of the original noise $\epsilon_{i,n}$. See also the related papers Chaker (2017) and Clinet and Potiron (2017, 2018c, 2018d). Here again the observation times are assumed to be regular, i.e. $\tau_{i,n} = iT/n$.

\smallskip
The authors assume that $\tilde{\epsilon}_{i,n}$ is with mean $0$, finite standard deviation and that $n \var [\tilde{\epsilon}_{i,n}] \rightarrow v$, which in turn implies that $\tilde{\epsilon}_{i,n} = O_p ( 1/\sqrt{n})$. To embed this assumption in our LPM framework, there is no harm assuming that 
$$\tilde{\epsilon}_{i,n} = n^{-\alpha} v^{\frac{1}{2}} \gamma_{\tau_{i,n}},$$ 
where $\alpha \geq 1/2$ and $\gamma_t$ are IID with null-mean and unity variance. They estimate $\nu$ and the underlying price as 
\begin{eqnarray*}
\widehat{\nu} & = & \underset{\nu}{\text{arg min}} \frac{1}{2} \sum_{i=1}^{N_n} ((Z_{\tau_{i,n},n} - Z_{\tau_{i-1,n},n}) - (h (I_{i,n},\nu) - h (I_{i-1,n},\nu)))^2,\\  \widehat{X}_{\tau_{i,n}} & = & Z_{\tau_{i,n},n} - h(I_{i,n}, \hat{\nu}).
\end{eqnarray*}
The authors then estimate the integrated volatility with 
$$ERV_{ext} = \sum_{i=1}^{N_n} ( \Delta \widehat{X}_{\tau_{i,n}})^2 + 2 \sum_{i=2}^{N_n}  \Delta \widehat{X}_{\tau_{i,n}} \Delta \widehat{X}_{\tau_{i-1,n}},$$
where $\Delta \widehat{X}_{\tau_{i,n}} = \widehat{X}_{\tau_{i,n}} - \widehat{X}_{\tau_{i-1,n}}$, and show the according central limit theory. Under suitable assumptions, they obtain the optimal convergence rate $n^{1/2}$ and the AVAR when $T=1$:
$$AVAR^{(ERV)} = 6\int_0^1 \sigma_t^4 dt + 8 v \int_0^1 \sigma_t^2 dt + 8 v^2.$$

\smallskip
They also consider another estimator (which they call E-QMLE) which consists in using the QMLE from Xiu (2010), which we considered as a local estimator in Example \ref{noise}, on the estimated observations $\widehat{X}_{\tau_{i,n}}$. They indicate that the E-QMLE might yield a smaller AVAR (see their discussion on p. 38), and they report in their numerical study that its finite sample performance is comparable to $ERV_{ext}$ (see Table 2 in p. 41). They do not investigate the corresponding central limit theory. 

\smallskip
With the theory provided in our paper, we cannot investigate the E-QMLE, but rather the E-(LPE of QMLE), i.e. we apply Example \ref{noise} on $\widehat{X}_{\tau_{i,n}}$. To keep notation of our paper, we denote $\widehat{\Theta}_n$ the E-(LPE of QMLE) estimator of volatility and $\widehat{\Theta}_n^{(BC)}$ its bias-corrected version (i.e. E-(BC LPE of QMLE)). The AVARs obtained in Theorem \ref{thEQMLE} are the same as in Theorem \ref{thQMLE}. This is due to the fact that the estimation of $\nu$ is very accurate featuring $n$ as a rate of convergence and thus the pre-estimation does not impact the AVAR. This was already the case for the $ERV_{ext}$ (see the proof of Theorem 3 on pp. 46-47 in Li et al. (2016)). 

\begin{figure}
\includegraphics[width=.6\linewidth]{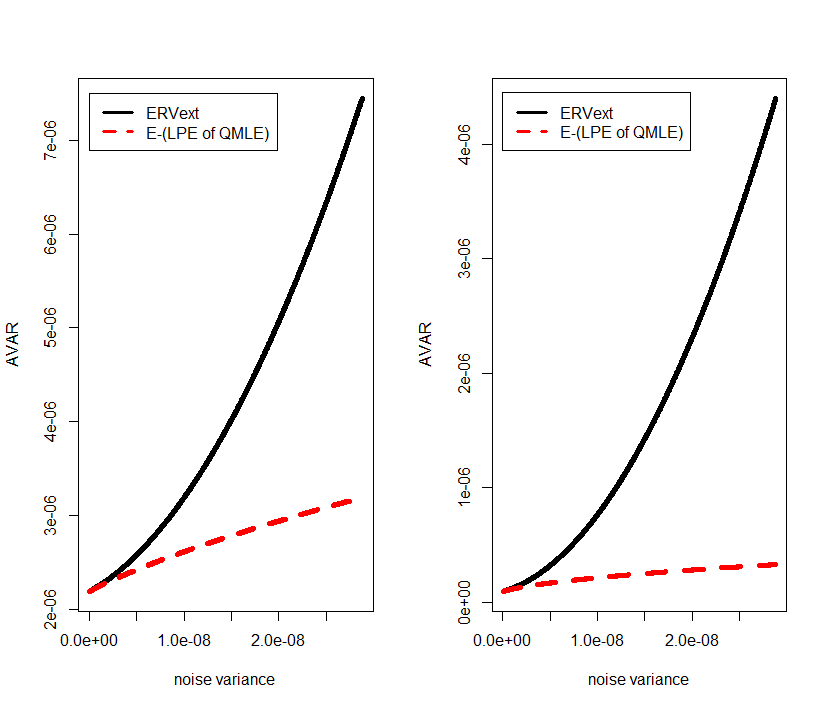}
\centering
\caption{AVAR of $ERV_{ext}$ and E-(LPE of QMLE) as a function of the noise variance, i.e. the variance of $\tilde{\epsilon}_{i,n}$. The horizon time is set to $T=1$ (which corresponds to 6.5 hours of intraday trading). On the left hand-side, we follow exactly the setting of the numerical study in Li et al. (2016), where $\sigma_t^2 = 0.000125$ if $0.05 \leq t < 0.95$ and $\sigma_t^2 = 15*0.000125$ otherwise. There is on average one observation a second, which corresponds to $n = 23,400$. On the right hand-side, the setting is the same except that we remove the jumps in volatility and consider $\sigma_t^2 = 0.000125$ for $0 \leq t \leq 1$.}
\label{AVAR}
\end{figure}

\smallskip
Recalling that the LPE of QMLE is conjectured to be more efficient than the QMLE, in particular this implies that E-(LPE of QMLE) is also conjectured to be more efficient than E-QMLE. In Figure \ref{AVAR}, we can see that E-(LPE of QMLE) highly improves the AVAR compared to the $ERV_{ext}$. The improvement gets bigger as the noise of $\tilde{\epsilon}_{i,n}$ increases. When setting the volatility and the noise variance as in the setting of the numerical study in Li et al. (2016), the ratio of AVARS is equal to $0.7$. When we further assume no jumps in volatility, this ratio goes to $0.2$. When choosing a bigger noise variance $1.44e-07$ which remains reasonable, this ratio is lower than $0.01$. The overall picture is clearly in favor of the E-(LPE of QMLE). We provide the theorem of this estimator in what follows.
\begin{thEQMLE} (E-(LPE of QMLE))
\label{thEQMLE}
 Under Assumption A in Li et al. (2016, p. 7):\\ (i) we assume that $\alpha > \frac{1}{2}$. Then, stably in law\footnote{Here and in the following statements, the stable convergence in law is with respect to the filtration considered in Li et al. (2016).}as $n \rightarrow \infty$,
\begin{eqnarray}
\label{thEQMLE1}
 n^{\frac{1}{2}} \Big( \widehat{\Theta}_n - T^{-1} \int_0^T \sigma_s^2 ds \Big) \rightarrow \Big(6 T^{-1} \int_0^T \sigma_s^4 ds \Big)^{\frac{1}{2}} \mathcal{N} (0,1).
\end{eqnarray}
(ii) when $\alpha = \frac{1}{2}$, we have stable convergence in law of 
\begin{eqnarray*}
   n^{\frac{1}{2}} \big( \widehat{\Theta}_n^{(BC)} - \Theta \big) \rightarrow \Big( T^{-1} \int_0^T \big(2 \sigma_s^4 + 4 \sigma_s^3 \sqrt{4v + \sigma_s^2} \big) ds \Big)^{\frac{1}{2}} \mathcal{N} (0,1).
\end{eqnarray*}
\end{thEQMLE}
We discuss now briefly how to estimate the higher powers of volatility, i.e. when $\theta_t^* = g(\sigma_t^2)$ with $g$ not being the identity function. We consider the estimators from Example \ref{expowers}. The difference with Example \ref{expowers} is that this estimator is used on the estimated price $\widehat{X}_{\tau_{i,n}}$ based on the information rather than on the raw price. The related theorem is given in what follows.
\begin{thEQMLEpowers} (powers of volatility)
\label{thEQMLEpowers}
Under Assumption A in Li et al. (2016, p. 7):\\ (i) We assume that $\alpha > \frac{1}{2}$. Then, stably in law as $n \rightarrow \infty$,
\begin{eqnarray}
 n^{\frac{1}{2}} \big( \widehat{\Theta}_n^{(BC,1)} - \Theta \big) \rightarrow \Big(6 T^{-1} \int_0^T \big( g' (\sigma_s^2) \big)^2 \sigma_s^4 ds \Big)^{\frac{1}{2}} \mathcal{N} (0,1).
\end{eqnarray}
(ii) When $\alpha = \frac{1}{2}$, we have
\begin{eqnarray*}
 n^{\frac{1}{2}} \big( \widehat{\Theta}_n^{(BC,2)} - \Theta \big) \rightarrow 
 \Big( T^{-1} \int_0^T (g' (\sigma_s^2))^2 \big(2 \sigma_s^4 + 4 \sigma_s^3 \sqrt{4v + \sigma_s^2} \big) ds \Big)^{\frac{1}{2}} \mathcal{N} (0,1).
\end{eqnarray*}
\end{thEQMLEpowers}

\subsection{Estimation of volatility using the model with uncertainty zones}
\label{p2uncertaintyzones}
 We introduce a time-varying friction parameter extension to the model with uncertainty zones introduced in Robert and Rosenbaum (2011). To incorporate microstructure noise in the model, we define $\alpha_n$ as the tick size, and the related asymptotics is such that $\alpha_n \rightarrow 0$. Correspondingly we assume that the observed price $Z_{\tau_{i,n},n}$ takes values on the tick grid (i.e. modulo of size $\alpha_n$). 
 
 \smallskip
 We discuss first a simple version of the model with uncertainty zones, which features endogeneity in arrival times. In a frictionless market, we can assume that all the returns (which we recall to be defined as $R_{i,n} = Z_{\tau_{i,n},n} - Z_{\tau_{i-1,n},n}$) have a magnitude of exactly one tick, and that the next transaction will occur when the latent price process crosses the mid-tick value $X_{\tau_{i,n}} + \frac{\alpha_n}{2}$ in case of the price goes up (or $X_{\tau_{i,n}} - \frac{\alpha_n}{2}$ when the price goes down). We extend this toy model in what follows.

\smallskip
The authors introduce the discrete variables $L_{i,n}$ that stands for the absolute size, in tick number, of the next return. In other words, the next observed price has the form $Z_{\tau_{i+1,n},n} = Z_{\tau_{i,n},n} \pm \alpha_n L_{i,n}$. They also introduce a continuous (possibly multidimensional) time-varying parameter $\chi_t$, and assume that conditional on the past, $L_{i,n}$ takes values on $\{1, \cdots, m \}$ with
$$\proba_{\tau_{i,n}} ( L_{i,n} = k) = p_k (\chi_{\tau_{i,n}})$$
for some unknown positive differentiable with bounded derivative functions $p_k$ such that $\sum_{k=1}^{m} p_k = 1$.

\smallskip
Also, the frictions induce that the transactions will not occur exactly when the efficient process crosses the mid-tick values. For this purpose, in the notation of Robert and Rosenbaum (2012), let $0 < \eta < 1$ be the parameter that quantifies the aversion to price change. The frictionless scenario corresponds to $\eta = 0$. Conversely, the agents are very averse to trade when $\eta$ is closer to $1$. If we define $X_t^{(\alpha)}$ as the value of $X_t$ rounded to the nearest multiple of $\alpha$, the sampling times are defined recursively as $\tau_{0,n} := 0$ and for any positive integer $i$ as
\begin{eqnarray}
\label{timesuncertainty} \tau_{i,n} := \inf \Big\{ t > \tau_{i-1,n} : X_t = X_{\tau_{i-1,n}}^{(\alpha_n)} - \alpha_n \big(L_{i-1,n} - \frac{1}{2} + \eta \big)\\ 
\text{ or } X_t = X_{\tau_{i-1,n}}^{(\alpha_n)} + \alpha_n \big(L_{i-1,n} - \frac{1}{2} + \eta \big) \Big\}. \nonumber
\end{eqnarray}
Correspondingly, the observed price is assumed to be equal to the rounded efficient price $Z_{\tau_{i,n},n} := X_{\tau_{i,n}}^{(\alpha_n)}$.

\smallskip
In the extension of (\ref{timesuncertainty}) when $\eta_t$ is time-varying, we assume that
the sampling times are defined recursively as $\tau_{i,n} := 0$ and for any positive integer $i$ as
$$\tau_{i,n} := \inf \Big\{ t > \tau_{i-1,n} : X_t = X_{\tau_{i-1,n}}^{(\alpha_n)} - \alpha_n \big(L_{i,n} - \frac{1}{2} + \eta_{\tau_{i-1,n}} \big)$$ 
\begin{eqnarray} \label{timevaryinguncertaintyzones}
\text{ or } X_t = X_{\tau_{i-1,n}}^{(\alpha_n)} + \alpha_n \big(L_{i,n} - \frac{1}{2} + \eta_{\tau_{i-1,n}} \big) \Big\}.
\end{eqnarray}
The idea behind the time-varying friction model with uncertainty zones is that we hold the parameter $\eta_t$ constant between two observations.

\smallskip
To express the model with uncertainty zones as a LPM, we consider that $\theta_t^* := (\sigma_t^2, \eta_t, \chi_t)$. Following the definition (p. 11) in Robert and Rosenbaum (2012), we further introduce a Brownian motion $W_t'$ independent of all the other quantities, and let $\Phi$ denote the cumulative distribution function of a standard Gaussian random variable. We specify the definition of $L_{i,n}$ related to $W_t'$ as 
\begin{eqnarray*}
g_{t,n} & = & \sup \{ \tau_{j,n} \text{ : } \tau_{j,n} < t \},\\
L_t' & = & \sum_{k=1}^m k \mathbf{1} \bigg\{ \Phi \Big( \frac{W_t' - W_{g_{t,n}}}{\sqrt{t - g_{t,n}}} \Big) \in \Big[ \sum_{j=1}^{k-1} p_j (\chi_t), \sum_{j=1}^k p_j (\chi_t) \Big] \bigg\}, 
\end{eqnarray*}
and $L_{i,n} = L_{\tau_i,n}'$. If we set the random innovation as the two-dimensional process $U_{i,n} := ((W_t - W_{\tau_{i-1,n}})_{t \geq \tau_{i-1,n}},$ $((W_t' - W_{\tau_{i-1,n}}')_{t \geq \tau_{i-1,n}})$, and the past as $P_{\tau_{i,n}} = (L_{i,n}, \text{sign} (R_{i,n}))$, we can deduce the form of $F_n$ in the model.\footnote{The advised reader will have noticed that a priori, $\text{sign} (R_{i,n})$ and $\eta_t$ are not independent, so that the assumptions of the LPM do not hold entirely. This problem can be circumvented as the former is actually conditionally independent from the latter.}

\smallskip
We provide in what follows the definition of the estimators. We are not interested in estimating directly $\chi_t$ and thus we consider the subparameter $\Theta := (\int_0^T \sigma_t^2 dt,$ $\int_0^T \eta_t dt)$ to be estimated. For $k=1, \cdots, m$ we define 
\begin{eqnarray*}
N_{t,k,n}^{(a)} & = & \sum_{i=1}^{N_n(t)} \mathbf{1}_{\{R_{i,n} R_{i-1,n} < 0 \text{ , } \mid R_{i,n} \mid = k \alpha_n \}},\\
N_{t,k,n}^{(c)} & = & \sum_{i=1}^{N_n(t)} \mathbf{1}_{\{R_{i,n} R_{i-1,n} > 0 \text{ , } \mid R_{i,n} \mid = k \alpha_n \}}
\end{eqnarray*}
as respectively the number of alternations and continuations of $k$ ticks. By alternation (continuation) of $k$ ticks, we mean that the return magnitude is of $k$ ticks with a direction opposite to (with the same direction as) the previous return. We define the estimator of $\eta$ as\footnote{Actually, the estimator considered here slightly differs from the original definition (p. 8) in Robert and Rosenbaum (2012) as it provides smaller theoretical finite sample bias. Asymptotically, both estimators are equivalent and thus all the theory provided in Robert and Rosenbaum (2012) can be used to prove Theorem \ref{cltuncertainty}.}
\begin{eqnarray} \label{hatetamod}
\widehat{\eta}_{t,n} := \sum_{k=1}^{m} \lambda_{t,k,n} u_{t,k,n},
\end{eqnarray}
with
\begin{eqnarray*}
\lambda_{t,k,n} & := & \frac{N_{t,k,n}^{(a)} + N_{t,k,n}^{(c)}}{\sum_{j=1}^{m} \big( 
N_{t,j,n}^{(a)} + N_{\alpha,t,j}^{(c)} \big) },\\
u_{t,k,n} & := & \max\bigg\{ 0, \min \bigg\{1, \frac{1}{2} \bigg(
k \bigg( \frac{N_{t,k,n}^{(c)}}{N_{t,k,n}^{(a)}} - 1 \bigg) + 1 \bigg) \bigg\} \bigg\},
\end{eqnarray*}
where $N_n(t)$ is defined as the integer satisfying $Z_{\tau_{N_n(t),n}} < t < Z_{\tau_{N_n(t)+1,n}}$, we assume that $C/0 := \infty$, and in particular $u_{\alpha,t,k} = 1$ when $N_{\alpha,t,k}^{(a)} = 0$. The key idea is that $u_{\alpha,t,k}$ are consistent estimators of $\eta$. 
Based on the friction parameter estimate, we can construct a consistent latent price estimator as
$$\widehat{X}_{\tau_{i,n}} = Z_{\tau_{i,n},n} - \alpha_n (1/2 - \widehat{\eta}_{t,n}) \text{sign} (R_{i,n}).$$
The estimator of integrated volatility is obtained using the usual realized volatility estimator on the estimated price defined as
$$\widehat{RV}_{t,n} = \sum_{i=1}^{N_n (t)} (\widehat{X}_{\tau_{i,n}} - \widehat{X}_{\tau_{i-1,n}})^2.$$
 The related local estimators $\widehat{\Theta}_{i,n} := (\widehat{\sigma}_{i,n}^2, \widehat{\eta}_{i,n})$ are constructed from local versions of $\big(\widehat{RV}_{t,n}, \widehat{\eta}_{t,n}\big)$.

\begin{cltexampleuncertainty} (Time-varying friction parameter model with uncertainty zones) \label{cltuncertainty}
 Let $\mathcal{G}_t$ be the filtration generated by $X_t$, $\chi_t$ and $\eta_t$. $\mathcal{G}_T$-stably in law as $n \rightarrow \infty$,
\begin{eqnarray}
 \label{main1} \alpha_n^{-1} \big( \widehat{\Theta}_n - \Theta \big) \rightarrow \Big( T^{-1} \int_0^T V_{\theta_s^*} ds \Big)^{\frac{1}{2}} \mathcal{N} (0,1),
\end{eqnarray}
where $V_{\theta}$ can be straightforwardly inferred from the definition of Lemma $4.19$ in p. 26 of Robert and Rosenbaum (2012).
\end{cltexampleuncertainty}
\begin{rkuncertaintycvrate}
(convergence rate) Note that, equivalently, the convergence rate in (\ref{main1}) is $n^{\frac{1}{2}}$ when $n$ corresponds to the expected number of observations. One can consult Remark 4 in Potiron and Mykland (2017) for more details about this. 
\end{rkuncertaintycvrate}

\subsection{Application in time series: the time-varying MA(1)}
\label{timeseries}
We first specify the LPM for a general one-dimensional time series. In that case, we assume that the observation times are regular. We further assume that the returns $R_{i,n}$ stand for time series observations. Finally, we assume that the time-varying time series can be expressed as the interpolation of $\theta_t^*$ via
\begin{eqnarray} \label{timevaringtimeseries}
 R_{i,n} = F_n \big( \{P_{s,n}\}_{0 \leq s \leq \tau_{i-1,n}}, U_{i,n}, \theta_{\tau_{i-1,n}}^* \big),
\end{eqnarray} 
where $\theta_t^*$ is assumed to be independent of all the innovations. When $\theta_t^*$ is constant, numerous time series\footnote{We can actually show that any time series in state space form can be expressed with a corresponding $F_n$ function.} are of the form (\ref{timevaringtimeseries}). 

\smallskip
We now discuss the specific MA(1) representation. Several time-varying extensions are possible and we choose to work with the time-varying parameter model 
$$R_{i,n}= \mu_{\tau_{i-1,n}} + \sqrt{\kappa_{\tau_{i-1,n}}} \lambda_{i,n} + \beta_{\tau_{i-1,n}} \sqrt{\kappa_{\tau_{i-1,n}}} \lambda_{i-1,n},$$ 
where $\lambda_{i,n}$ are standard normally-distributed white noise error terms, and $\kappa_{t}$ is the time-varying variance. The three-dimensional parameter is defined as $\theta_t^* := (\mu_t, \beta_t, \kappa_t) \in \reels^2 \times \reels_*^+$. We fix both the innovation and the past equal to the white noise $U_{i,n} = \lambda_{i,n}$ and $P_{\tau_{i,n},n} = \lambda_{i,n}$. We have thus expressed the MA(1) as a LPM.

\smallskip
We discuss how to estimate the parameters in what follows. For simplicity, we assume that $\mu_t = 0$. The target quantity is thus equal to $\Theta = ( \int_0^T \beta_t dt, \int_0^T \kappa_t dt)$. The local estimator is the MLE  (see Hamilton (1994), Section 5.4). On each block (of size $h_n$), the MLE bias is of order $h_n^{-1}$ (Tanaka (1984)) and thus the bias condition (\ref{E6110}) is not satisfied. Nonetheless, we can correct for the bias up to the order $O(h_n^{-2})$ as follows. We define the bias-corrected estimator as $$\widehat{\Theta}_{i,n}^{(BC)} = \widehat{\Theta}_{i,n} - b(\widehat{\Theta}_{i,n},h_n),$$ where the bias function $b(\theta,h)$ can be derived following the techniques in Tanaka (1984). In particular this implies that the bias-corrected estimator satisfies the bias condition if $h_n$ is chosen such that $n^{1/4} = o (h_n)$. In practice this bias can be obtained by Monte-Carlo simulations (see our simulation study).

\smallskip
In the parametric case and in a low frequency asymptotics where $T \rightarrow \infty$ and observations times are $0, \Delta, \cdots, T = n\Delta$ with $\Delta > 0$, known results (see, e.g., the proof of Proposition I in pp. 391-393 (A\"{i}t-Sahalia et al. (2005)) show that the asymptotic variance of the MLE is such that 
$$n^{1/2} \big((\widehat{\beta}, \widehat{\kappa}) - (\beta, \kappa) \big) \rightarrow  \left(\begin{matrix} 1-\beta^2 & 0\\ 
0 & 2 \kappa^2
\end{matrix}\right)^{1/2} \mathcal{N} (0,1).$$
The following theorem provides the time-varying version of the asymptotic theory when T is fixed.

\begin{exMA1} (Time-varying MA(1))
\label{exMA1}
 Let $\mathcal{F}_t^{\theta}$ the filtration generated by $\theta_t^*$. We assume that $n^{1/4} = o (h_n)$ and Condition (P2). Then, $\mathcal{F}_T^{\theta}$-stably in law as $n \rightarrow \infty$,
\begin{eqnarray*}
 n^{\frac{1}{2}} \big( \widehat{\Theta}_n^{(BC)} - \Theta \big) \rightarrow \Bigg( T^{-1} \left(\begin{matrix} \int_0^T (1-\beta_s^2) ds & 0\\ 
0 & \int_0^T 2 \kappa_s^2 ds
\end{matrix}\right) \Bigg)^{\frac{1}{2}} \mathcal{N} (0,1).
\end{eqnarray*}
\end{exMA1}

\subsection{Further examples}
Two further examples include our own recent work. Potiron and Mykland (2017) introduced a bias-corrected Hayashi-Yoshida estimator (Hayashi and Yoshida (2005)) of the high-frequency covariance and showed the corresponding CLT under endogenous and asynchronous observations. To model duration data, Clinet and Potiron (2018a) built a time-varying Hawkes self-exciting process, derived the bias-corrected MLE and showed the CLT of the corresponding LPE.

\subsection{Discussion}
We provide in what follows a discussion on the efficiency and robustness of the specific examples considered in this section. The subsequent techniques may also be useful to tackle other examples from the literature.
\subsubsection{Efficiency}
There are many problems where $n^{1/2}$ is rate-optimal from Gloter and Jacod (2001), such as all the examples considered in this section. In addition, the local feature of the technology should yield efficiency in case the parametric estimator is efficient itself. This is the case of (\ref{thQMLE2}) in Example \ref{noise}, Theorem \ref{thQMLEpowers} (ii) in Example \ref{expowers}, Theorem \ref{thEQMLE} (ii) and Theorem \ref{thEQMLEpowers} (ii) in Example \ref{exyingying}, Theorem \ref{exMA1} in Example \ref{timeseries}, where the parametric estimator achieves the Cram\'{e}r-Rao bound of efficiency locally. 

\smallskip
In the case of (\ref{thQMLE1}) in Example \ref{noise}, i.e. when estimating volatility assuming that the noise is very small $\epsilon_{i,n} = o_p ( 1/\sqrt{n})$, the asymptotic variance is equal to $6 T^{-1} \int_0^T \sigma_s^4 ds$, whereas the efficient bound $2 T^{-1} \int_0^T \sigma_s^4 ds$ is attained by the RV. This increases the variance by a factor of 3, which is also observed on the MLE (when assuming the volatility is constant) when misspecified on a model which does not incorporate microstructure noise (see, e.g., Section 2.4 pp. 1486-1487 in Barndorff-Nielsen et al. (2008)).

\subsubsection{Robustness to drift and jumps in the latent price process}
We focus on the specific case where the observations are related to a latent continuous-It\^{o} price model $dX_t = \int_0^t \sigma_u dW_u$, as in
 Example \ref{noise}-\ref{p2uncertaintyzones} (Example \ref{timeseries} considers a time series without any underlying price process). We discuss how we can add a drift and jumps in $X_t$ in those examples.
 
 \smallskip
 We first show how to add a drift component. By Girsanov theorem, in conjunction with local arguments (see, e.g., pp. 158-161 in Mykland and Zhang (2012)), we can weaken the price and volatility local-martingale assumption by allowing them to follow an It\^{o}-process (of dimension 2 in case of volatility or powers of volatility estimation), 
 with a volatility matrix locally bounded and locally bounded away from 0, and drift which is also locally bounded.
 
 \smallskip
It is also easy to see that we can allow for finite activity jumps in $X_t$. To do that, we assume that $\widehat{\Theta}_{i,n}$ is taking values on a compact set\footnote{The MLE is always performed on a compact set, so the assumption is trivially satisfied in that case, which corresponds to Example \ref{noise}-\ref{exyingying}. Moreover, the estimator of $\eta$ in Example \ref{p2uncertaintyzones} is bounded by definition, but one would need to bound the volatility estimator to apply the technique.}. Consider $J_n \subset \{1, \cdots, B_n \}$ the set of blocks where there is at least one jump in $X_t$. As the number of blocks $B_n \rightarrow \infty$, the cardinality of $J_n$ is at most finite, and thus we have that 
$$\frac{1}{T} \sum_{i=1}^{N_n} \widetilde{\Theta}_{i,n} \Delta \Tau_{i,n} \approx \frac{1}{T} \sum_{i \notin J_n} \widetilde{\Theta}_{i,n} \Delta \Tau_{i,n}.$$
It is then immediate to adapt the proof of the CLT. On the other hand, if infinitely many jumps are possible in both the price process and the parameter, the theoretical development is beyond the scope of this paper.

\subsubsection{Robustness to jumps in $\theta_t^*$}
By a similar reasoning as for when adding jumps in $X_t$, the techniques of this paper are robust to jumps (of finite activity) in $\theta_{t}^{*}$ in all our examples.

\subsubsection{Non regular observation times}
\smallskip
We also assume here that there is a latent price process and reason about the type of observation times which falls into the LPM. We consider first the time deformation of Barndorff-Nielsen et al. (2008, Section 5.3, pp. 1505-1507). To express their setting as a LPM, we assume that the observation times are of the form 
\begin{eqnarray}
\label{endotimes}
\tau_{i,n} = \Gamma_{i/(nT)},
\end{eqnarray}
where $\Gamma_t$ is a stochastic process satisfying $\Gamma_t = \int_0^t \tilde{\Gamma}_u^2 du$, with $\tilde{\Gamma}_t$ a strictly positive parameter of the LPM. We can then construct a (change of time) process $\tilde{X}_t = X_{\Gamma_t}$ so that for $\tilde{X}_t$ the observations are regular. In view of Dambis Dubins-Schwarz theorem (see, e.g., Theorem 1.6 on p. 181 in Revuz and Yor (1999)) we have that $[X]_{T}  = [\tilde{X}]_{\Gamma_{T}}.$ In addition, it is immediate to see that Condition (T) and (\ref{nonregeq2}) hold in that case. 

\smallskip
Alternatively one can assume that the quadratic variation of time (see, e.g., Assumption A on p. 1939 in Mykland and Zhang (2006)) exists and that observation times are independent of the price process. Under suitable assumptions, we can also show that Condition (T) and (\ref{nonregeq2}) hold.

\smallskip
Our setting can actually allow for (some kind of) endogenous stopping times as in the case of the model with uncertainty zones detailed in Example \ref{p2uncertaintyzones}. The type of endogeneity is such that there is no asymptotic bias in the related central limit theorem. 

\smallskip
Finally, the model allows for endogenous observation times in the general multidimensional HBT model introduced in Potiron and Mykland (2017). In that case, the central limit theorem features an asymptotic bias.\footnote{Details about the model can be found in a previous version of the manuscript circulated under the name "Estimating the Integrated Parameter of the Locally Parametric Model in High-Frequency Data".}

\subsubsection{Estimating time-varying functions of $\theta_t^*$}
Another nice corollary about the introduced theory is that we can obtain the central limit theorem of the powers of the integrated parameter $g (t, \theta_t^*)$ for $g$ smooth enough when using the local estimates $g (\Tau_{i-1,n}, \widehat{\Theta}_{i,n})$. Essentially, the proof uses on each block a Taylor expansion as in the delta method. We apply the technique on the local QMLE in Example \ref{expowers} and on an adapted estimator from Li et al. (2016) in Example \ref{exyingying} to estimate the higher powers of volatility. 

\section{Numerical study: estimation of volatility with the QMLE}
\label{numericalstudy}
\subsection{Goal of the study}
To investigate the finite sample performance of the LPE, we consider the local QMLE introduced in Section \ref{noise}. The goal of the study is twofold. First, we want to investigate how the LPE performs compared to the global QMLE. Second, we want to discuss about the choice of the number of blocks $B_n$ in practice. Complementary simulation results can be found in Clinet and Potiron (2018b).

\subsection{Model design}
We perform Monte-Carlo simulations of M=1,000 days of high-frequency observations where the related horizon time is set to $T = 1/252$ (i.e. annualized). One working day stands for 6.5 hours of trading activity, which can also be expressed as 23,400 seconds. We consider three high-frequency sampling frequency scenarios: every second, every other second, and every three seconds. 

\smallskip
We perform local QMLE with number of blocks ranging from $B_n=1$ (i.e. the global QMLE case) to $B_n=20$. The corresponding number of observations per block ranges from $h_n=1,170$ to $h_n = 23,400$ in the case of 1-second sampling frequency, from $h_n=585$ to $h_n = 11,700$ if we sample ever other second, and from $h_n=390$ to $h_n = 7,800$ when subsampling every three seconds. Note that the minimal number of observations per block remains reasonable in view of the finite sample performance of the global QMLE (see the numerical study in Xiu (2010)).

\smallskip
We bring forward the Heston model with U-shape intraday seasonality component and jumps in volatility as
\begin{eqnarray*}
dX_t &=& b dt + \sigma_{t}dW_t,\\
\sigma_t &=& \sigma_{t-,U} \sigma_{t,SV},
\end{eqnarray*}
where
\begin{eqnarray*} 
\sigma_{t,U} & = & C + Ae^{-at/T} + De^{-c(1-t/T)} -  \beta\sigma_{\tau-,U}\mathbf{1}_{\{t \geq \tau \}},\\
d\sigma_{t,SV}^2 & = & \alpha(\bar{\sigma}^2 - \sigma_{t,SV}^2)dt + \delta \sigma_{t,SV}d\bar{W}_t, 
\end{eqnarray*}
where the parameters are set to $b = 0.03$, $C=0.75$, $A=0.25$, $D=0.89$, $a=10$, $c=10$, the volatility jump size parameter $\beta=0.5$, the volatility jump time $\tau$ follows a uniform distribution on $[0,T]$, $\alpha = 5$, $\bar{\sigma}^2 = 0.1$, $\delta = 0.4$, $\bar{W}_t$ is a standard Brownian motion such that  $d\langle W,\bar{W} \rangle_t = \overline{\phi} dt$, $\overline{\phi} = -0.75$, $\sigma_{0,SV}^2$ is sampled from a Gamma distribution of parameters $(2\alpha\bar{\sigma}^2/\delta^2,\delta^2/2\alpha)$, which corresponds to the stationary distribution of the CIR process. For further reference, see Clinet and Potiron (2018b). The model is almost the same as that of Andersen et al. (2012). Finally, the noise is assumed normally distributed with zero-mean and constant variance $v$ set so that the noise to signal ratio defined as
\begin{eqnarray}
\xi^2 = \frac{a_0^2}{\sqrt{T\int_0^T \sigma_u^4 du} }
\end{eqnarray}
is equal to $\xi^2 = 0.0001$.

\subsection{Results}
 Table \ref{tablenum} reports the sample bias, standard deviation and the RMSE of the local quasi maximum likelihood volatility estimator. The number of blocks ranges from $B_n=1$, which corresponds to the global QMLE, to $B_n=20$. Regardless of the sampling frequency, the numerical experiment results are quite similar. There is a very small sample bias (the bias to standard deviation ratio magnitude is around $0.03$), which increases with the number of blocks while staying very small, all of which hinting that the it is not necessary to use a bias correction of the local QMLE in practice. The standard deviation decreases and then stays (roughly) stable. The picture for the RMSE is the same, all of this very much in line with the fact that almost all the theoretical gain is already obtained in the case of $B=8$ blocks (see Clinet and Potiron (2018b)). Finally, the smallest RMSE is obtained with $B_n=19$ blocks when sampling at 1-second frequency, $B_n=8$ in case of 2-second frequency and $B_n=14$ with 3-second subsampling observations indicating that the finer the sampling frequency the larger the number of blocks should be used. The gains in terms of RMSE goes almost up to $10\%$ when sampling at the finest frequency, whereas less than $5\%$ in the other scenarios.
 \begin{table}[p]
 \centering
  \begin{tabular}{| c | c | c | c | c | c | c | c | c | c |}
   \hline
    Samp.freq.& 1 sec. & 1 sec. & 1 sec. & 2 sec. & 2 sec. & 2 sec. & 3 sec. & 3 sec. & 3 sec. \\
   \hline
    Nb.blocks & bias & s.d. & RMSE & bias & s.d. & RMSE & bias & s.d. & RMSE\\
  \hline
    1 & -2.398 & 8.158 & 8.162 & -2.503 & 10.813 & 10.814& -0.492 & 11.798 & 11.798 \\
    2 & -2.614 & 7.938 & 7.943 & -3.604 & 10.634 & 10.640& -0.700 & 11.642 & 11.642 \\
   3 & -2.882 & 7.820 & 7.825 & -4.041 & 10.537 & 10.544& -0.600 & 11.615 & 11.615 \\
    4 & -2.864 & 7.717 & 7.722 & -4.295 & 10.500 & 10.508& -1.210 & 11.596 & 11.597 \\
     5 & -3.181 & 7.720 & 7.727 & -4.757 & 10.528 & 10.539& -1.882 & 11.587 & 11.589 \\
    6 & -3.396 & 7.695 & 7.702 & -4.918 & 10.502 & 10.514& -2.213 & 11.610 & 11.612 \\
   7 & -3.662 & 7.665 & 7.674 & -5.373 & 10.523 & 10.537& -2.919 & 11.567 & 11.571 \\
    8 & -3.561 & 7.636 & 7.645 & -5.561 & 10.474 & 10.489& -3.388 & 11.601 & 11.606 \\
     9 & -4.225 & 7.636 & 7.648 & -6.344 & 10.557 & 10.576& -3.372 & 11.571 & 11.576 \\
    10 & -4.029 & 7.657 & 7.668 & -6.646 & 10.536 & 10.557& -4.400 & 11.613 & 11.621 \\
   11 & -4.503 & 7.593 & 7.607 & -6.876 & 10.526 & 10.548& -5.072 & 11.638 & 11.649 \\
    12 & -4.558 & 7.634 & 7.648 & -7.495 & 10.522 & 10.549& -5.580 & 11.629 & 11.642 \\
     13 & -4.769 & 7.644 & 7.659 & -8.045 & 10.548 & 10.578& -6.485 & 11.618 & 11.636 \\
    14 & -5.058 & 7.643 & 7.660 & -8.340 & 10.495 & 10.529& -7.282 & 11.533 & 11.555 \\
   15 & -5.416 & 7.591 & 7.610 & -8.394 & 10.498 & 10.531& -7.589 & 11.680 & 11.704 \\
    16 & -5.288 & 7.610 & 7.629 & -8.752 & 10.491 & 10.527& -8.452 & 11.607 & 11.638 \\
     17 & -5.638 & 7.608 & 7.629 & -8.856 & 10.457 & 10.494& -8.963 & 11.619 & 11.653 \\
    18 & -5.843 & 7.604 & 7.626 & -10.093 & 10.517 & 10.564& -9.239 & 11.625 & 11.661 \\
   19 & -6.283 & 7.568 & 7.594 & -10.270 & 10.499 & 10.549& -10.611 & 11.658 & 11.706 \\
    20 & -6.109 & 7.644 & 7.668 & -10.488 & 10.568 & 10.620& -10.644 & 11.603 & 11.652 \\
    
   \hline
  \end{tabular}
  \caption{In this table, we report the sample bias ($\times 10^7$), the standard deviation ($\times 10^6$) and the RMSE ($\times 10^6$) for local QMLE with number of blocks ranging from $B_n=1$ (i.e. the global QMLE case) to $B_n=20$. The number of seconds for one working day is $23,400$. The number of Monte-carlo simulations is 1,000. Three sampling frequencies are considered: every second, every other second, and every three seconds.}
\label{tablenum}
 \end{table}

\section{Conclusion}
In this paper, we have introduced a general framework to provide theoretical tools to build central limit theorems of convergence rate $n^{1/2}$ in a time-varying parameter model. We have applied successfully the method to investigate estimation of volatility (possibly under trading information), higher powers of volatility, the time-varying parameters of the model with uncertainty zones and the MA(1). This allowed us to obtain estimators robust to time-varying quantities, more efficient and/or new estimators of quantities (such as in the case of higher powers of volatility).

\smallskip
Subsequently, we believe that many other examples can be solved using the framework of our paper, which is simple and natural. This was successfully done in our related papers Potiron and Mykland (2017) and Clinet and Potiron (2018a). In those instances, the regular conditional distribution trick significantly simplified the work of the proofs.

\bibliographystyle{plain}

\begin{thebibliography}{186}
  
\bibitem{aitsahaliaal2017}
  A\"{i}t-Sahalia, Y., J. Fan, R. Laeven, C.D. Wang and X. Yang (2017).
  Estimation of the continuous and discontinuous leverage effects. \emph{Journal of the American Statistical Association}, 1-15
  

\bibitem{aitsahalia2010}
 A\"{i}t-Sahalia, Y., J. Fan and D. Xiu (2010). 
 High-frequency covariance estimates with noisy and asynchronous financial data. 
 \emph{Journal of the American Statistical Association} 105(492), 1504-1517.
  
 \bibitem{aitsahaliaal2005}
  A\"{i}t-Sahalia, Y., P.A. Mykland and L. Zhang (2005).
  How often to sample a continuous-time process in the presence of market microstructure noise.
  \emph{Review of Financial Studies} 18, 351-416.
  
   \bibitem{aitsahaliaal2016}
  A\"{i}t-Sahalia, Y. and D. Xiu (2016).
  A Hausman test for the presence of
market microstructure noise in high frequency data. To appear in \emph{Journal of Econometrics}.

\bibitem{altmeyer2015}
Altmeyer, R. and M. Bibinger (2015). Functional stable limit theorems for quasi-efficient spectral covolatility estimators. 
\emph{Stochastic Processes and their Applications} 125(12), 4556-4600.

\bibitem{andersen2012}
   Andersen, T. G., D. Dobrev and E. Schaumburg (2012). Jump-robust volatility estimation using nearest neighbor truncation. \emph{Journal of Econometrics}, 169:75-93.

  \bibitem{andersen2014}
  Andersen, T.G., D. Dobrislav and E. Schaumburg (2014).
  A robust neighborhood truncation approach to estimation of integrated quarticity,
  \emph{Econometric Theory} 30(1), 3-59.
  
  \bibitem{barndorff2002}	
Barndorff-Nielsen, O.E. and N. Shephard (2002). Econometric analysis of realized volatility and its use in estimating stochastic volatility models. \emph{Journal of the Royal Statistical Society: Series B (Statistical Methodology)} 64(2), 253-280.

\bibitem{Barndorff2006}
Barndorff-Nielsen, O.E., S.E. Graversen, J. Jacod, M. Podolskij and N. Shephard (2006). 
A central limit theorem for realised power and bipower variations of continuous semimartingales. 
\emph{From stochastic calculus to mathematical finance}. Springer Berlin Heidelberg, 33-68.

\bibitem{barndorff2008}
Barndorff-Nielsen, O.E., P.R. Hansen, A. Lunde, and N. Shephard (2008). 
Designing realized kernels to measure the ex post variation of equity prices in the presence of noise. 
\emph{Econometrica} 76(6), 1481-1536.

\bibitem{breiman}
Breiman, L. (1992)
\emph{Probability}. 
Classics in Applied Mathematics 7, Society for Industrial and Applied Mathematics (SIAM), Philadelphia, PA. 

\bibitem{chaker2017}
Chaker, S. (2017). On high frequency estimation of the frictionless price: The use of observed liquidity variables. \emph{Journal of Econometrics} 201, 127-143.

\bibitem{clinetpotiron2017}
Clinet, S. and Y. Potiron (2017). \emph{Estimation for high-frequency data under parametric market microstructure noise}. Working paper available at Arxiv 1712.01479.

\bibitem{clinetpotiron2018a}
Clinet, S. and Y. Potiron (2018a). 
Statistical Inference for the Doubly Stochastic Self-exciting Process. 
\emph{Bernoulli} 24(4B), 3469-3493. 

\bibitem{clinetpotiron2018b}
Clinet, S. and Y. Potiron (2018b). Efficient asymptotic variance reduction when estimating volatility in high frequency data. \emph{Journal of Econometrics} 206, 103-142.

\bibitem{clinetpotiron2018c}
Clinet, S. and Y. Potiron (2018c). \emph{Testing if the market microstructure noise is fully explained by the informational content of some variables from the limit order book}. Working paper available at Arxiv 1709.02502.

\bibitem{clinetpotiron2018d}
Clinet, S. and Y. Potiron (2018d). \emph{A relation between the efficient, transaction and mid prices: Disentangling sources of high frequency market microstructure noise}. Working paper available at SSRN 3167014.


\bibitem{dahlhaus1997}
Dahlhaus, R. (1997).
Fitting time series models to nonstationary processes,
\emph{Annals of Statistics} 25(1), 1-37.

\bibitem{dahlhaus2000}
Dahlhaus, R. (2000).
A likelihood approximation for locally stationary processes,
\emph{Annals of Statistics} 28(6), 1762-1794.

\bibitem{dahlhausrao2006}
Dahlhaus, R. and S.S. Rao (2006).
Statistical inference for time-varying ARCH processes,
\emph{Annals of Statistics} 34(3), 1075-1114.

\bibitem{daxiu2017}
Da, R. and D. Xiu (2017).
\emph{When Moving-Average Models Meet High-Frequency Data:
Uniform Inference on Volatility}. Working paper
available on Dacheng Xiu's website.

\bibitem{fangijbels1996}
Fan, J. and I. Gijbels (1996). 
\emph{Local polynomial modelling and its applications: monographs on statistics and applied probability 66}. CRC Press.

\bibitem{fanzhang1999}
Fan, J. and W. Zhang (1999). 
Statistical estimation in varying coefficient models. 
\emph{Annals of Statistics}, 1491-1518.

\bibitem{gloterjacod2001I}
Gloter, A. and J. Jacod (2001). 
Diffusions with measurement errors. I. Local asymptotic normality. 
\emph{ESAIM: Probability and Statistics} 5, 225-242.

\bibitem{giraitis2014}
Giraitis, L., G.Kapetanios and T. Yates (2014). 
Inference on stochastic time-varying coefficient models. 
\emph{Journal of Econometrics} 179, 46-65.

\bibitem{hallheyde1980}
Hall, P. and C.C. Heyde (1980). 
\emph{Martingale Limit Theory and its Application}. Academic
Press, Boston.

\bibitem{hamilton1994}
Hamilton, J.D. (1994). 
\emph{Time series analysis}. 
Vol. 2. Princeton: Princeton university press.

\bibitem{hastietibhirani1993}
Hastie T. and R. Tibshirani (1993). 
Varying-coefficient models. 
\emph{Journal of the Royal Statistical Society: Series B (Statistical Methodology)} 757-796.

\bibitem{hayashiyoshida2005}
Hayashi, T. and N. Yoshida (2005). 
On covariance estimation of non-synchronously observed diffusion processes. 
\emph{Bernoulli} 11, 359-379.

\bibitem{jacod1997}
Jacod, J. (1997). On continuous conditional Gaussian martingales and stable convergence in law. \emph{S\'eminaire de Probabilit\'es XXXI}, 232-246.

\bibitem{jacod2010}
Jacod, J., M. Podolskij and M. Vetter (2010). Limit theorems for moving averages of discretized processes plus noise. \emph{Annals of Statistics} 38(3), 1478-1545.

\bibitem{jacodprotter1998} 
Jacod, J. and P. Protter (1998). 
Asymptotic error distributions for the Euler method for stochastic differential equations. 
\emph{Annals of Probability} 26, 267-307.

\bibitem{jacodprotter2011}
Jacod, J. and P. Protter (2011). 
\emph{Discretization of Processes}. 
Springer.

\bibitem{jacodrosenbaum2013}
Jacod, J. and M. Rosenbaum (2013). 
Quarticity and other functionals of volatility: efficient estimation. 
\emph{Annals of Statistics} 41(3), 1462-1484.

\bibitem{jacodshiryaev2003}
Jacod, J. and A. Shiryaev (2003). 
\emph{Limit Theorems For Stochastic Processes} (2nd ed.). 
Berlin: Springer-Verlag.

\bibitem{kimnelson2006}
Kim, C. J. and Nelson, C. R. (2006). 
Estimation of a forward-looking monetary policy rule: A time-varying parameter model using ex post data. 
\emph{Journal of Monetary Economics} 53(8), 1949-1966.

\bibitem{kristensen2010}
Kristensen, D. (2010). Nonparametric filtering of the realized spot volatility: A kernel-based approach. 
\emph{Econometric Theory} 26(01), 60-93.

\bibitem{li2016}
Li, Y., S. Xie and X. Zheng (2016). Efficient estimation of integrated volatility incorporating trading information. 
\emph{Journal of Econometrics} 195(1), 33-50.

\bibitem{mancino2012}
Mancino, M.E. and S. Sanfelici (2012).
Estimation of quarticity with high-frequency data.
\emph{Quantitative Finance} 12(4),607-622.

\bibitem{myklandzhang2006}
Mykland, P.A., and L. Zhang (2006) 
ANOVA for Diffusions and It\^{o} Processes. 
\emph{Annals of Statistics} 34, 1931-1963.
  
  \bibitem{myklandzhang2009}
Mykland, P.A. and L. Zhang (2009). 
Inference for Continuous Semimartingales Observed at High Frequency. 
\emph{Econometrica} 77, 1403-1445.

\bibitem{myklandzhang2011}
Mykland, P.A. and L. Zhang (2011). 
The Double Gaussian Approximation for High Frequency Data. 
\emph{Scandinavian Journal Statistics} 38, 215-236.

\bibitem{myklandzhang2012}
Mykland, P.A. and L. Zhang (2012). 
The econometrics of High Frequency Data. 
In M. Kessler, A. Lindner and M. S\o rensen (eds.), 
\emph{Statistical Methods for Stochastic Differential Equations}, 109-190. Chapman nad Hall/CRC Press.

  \bibitem{myklandzhang2017}
  Mykland, P.A. and L. Zhang (2017).
  Assessment of Uncertainty in High Frequency Data: The Observed Asymptotic Variance \emph{Econometrica} 85(1), 197-231.

\bibitem{potironmykland2017}
Potiron, Y. and P.A. Mykland (2017). 
Estimation of integrated quadratic covariation with endogenous sampling times. \emph{Journal of Econometrics} 197, 20-41. 

\bibitem{podolskij2010}
Podolskij, M. and M. Vetter (2010). 
Understanding limit theorems for semimartingales: a short survey. 
\emph{Statistica Neerlandica} 64(3), 329-351.

\bibitem{reiss2011}
Rei\ss, M. (2011). 
Asymptotic equivalence for inference on the volatility from noisy observations. 
\emph{Annals of Statistics} 39(2), 772-802.

\bibitem{renault2012}
Renault, E., C. Sarisoy and B.J.M. Werker (2017). Efficient estimation of integrated volatility and related processes.
\emph{Econometric Theory} 33(2), 439-478.

\bibitem{revuzyor}
Revuz, D. and M. Yor (1999). 
\emph{Continuous Martingales and Brownian motion}. 
3rd ed., Germany: Springer.
  
  \bibitem{robert2011} 
Robert, C.Y. and M. Rosenbaum (2011). A new approach for the dynamics of ultra-high-frequency data: the model with uncertainty zones. 
\emph{Journal of Financial Econometrics} 9, 344-366. 

\bibitem{robert2012} 
Robert, C.Y. and M. Rosenbaum (2012). Volatility and covariation estimation when microstructure noise and trading times are endogenous. 
\emph{Mathematical Finance} 22(1), 133-164.

\bibitem{stockwatson1998}
Stock, J. H. and M. W. Watson (1998). 
Median unbiased estimation of coefficient variance in a time-varying parameter model. 
\emph{Journal of the American Statistical Association} 93(441), 349-358.

\bibitem{tanaka1984}
Tanaka K. (1984). 
An asymptotic expansion associated with the maximum likelihood estimators in ARMA models. 
\emph{Journal of the Royal Statistical Society: Series B (Statistical Methodology)} 58-67.

\bibitem{wangmykland2014}
Wang, C.D. and P.A. Mykland (2014).
The Estimation of Leverage Effect with High Frequency Data.
\emph{Journal of the American Statistical Association} 109, 197-215.

\bibitem{xiu2010}
Xiu, D. (2010).
Quasi-maximum likelihood estimation of volatility with high frequency data.
\emph{Journal of Econometrics} 159, 235-250.
\end{thebibliography}

\pagebreak

\pagebreak

\pagebreak
\section*{APPENDIX}
\section{Consistency in a simple model} 
\label{cons}
The purpose of this section is to provide an outline of the LPM and the conditions of the CLT by investigating the simpler problem of consistency in the case of a simple model. Two toy examples, the estimation of volatility with regular not noisy observations, and the estimation of the rate of a Poisson process, are discussed extensively. Morever, techniques of proofs are also mentioned throughout the section. The obtained conditions are illustrative. Proofs of the conditions along with proofs that conditions hold in the two toy examples can be found in Section \ref{proofs}. Finally, some detailed mathematical definitions can be found in what follows too.
\subsection*{The simple model} 
We focus on a simple setting in this section. First, we work with one dimensional returns, i.e. $d := 1$. Also, we assume that the observations are regular, so that $\tau_{i,n} = \frac{i}{n} T$.  The parametric model is assumed to be very simple, in particular there is no past dependence in the returns. It assumes that there exists a parameter $\theta^* \in K$ such that $R_{i,n}$ are independent and 
identically distributed (IID) random functions of $\theta^*$. If we introduce $U_{i,n}$ an adequate IID sequence of random variables with distribution $U$ which can depend on $n$, we can express the returns as
\begin{eqnarray} \label{paramret}
R_{i,n} & := & F_n \big( U_{i,n}, \theta^* \big),
\end{eqnarray}
where $F_n(x,y)$ is a non-random function. In (\ref{paramret}), $U_{i,n}$ can be seen as the \emph{random innovation}. 

\smallskip
Since $\theta_t^*$ can in fact be time-varying, $R_{i,n}$ do not necessarily follow (\ref{paramret}) in the time-varying parameter model. A formal time-varying generalization of (\ref{paramret}) will be given in (\ref{FUT1}). In general, $R_{i,n}$ are neither identically distributed nor independent. $R_{i,n}$
are not even necessarily conditionally independent given the true parameter process 
$\theta_t^*$, as we can see in the following two toy examples.

\begin{notIID} \label{estimatingvolatility}
(estimating volatility) Consider when $\theta_t^* := \sigma_t^2$ (the volatility is thus assumed 
to follow (\ref{paramito})), and $R_{i,n} := 
\int_{\tau_{i-1,n}}^{\tau_{i,n}} \sigma_s dW_s$, where $W_t$ is a standard $1$-dimensional Brownian motion. In this case, the parameter space is $K := \reels_{*}^{+}$. The parametric model assumes $\theta^* := \sigma^2$ and that the distribution of the returns is 
$R_{i,n} := \sigma \Delta W_{\tau_{i,n}}$, where $\Delta W_{\tau_{i,n}} := W_{\tau_{i,n}} - W_{\tau_{i-1,n}}$ is the increment of the Brownian motion 
between the $(i-1)$th observation time and the $i$th observation time and $\sigma^2$ is the fixed volatility. Under that assumption, the returns are IID. Under the time-varying parameter model, $R_{i,n}$ are clearly not necessarily IID, and they are also not necessarily conditionally independent given 
the whole volatility process $\sigma_t^2$ if there is a leverage effect.
\end{notIID}

\begin{notIID2} \label{estimatingratepoisson}
(estimating the rate of a Poisson process) Suppose the statistician observes data on the number of events (such as trades) in 
an arbitrary asset, and thinks the number of events happening between $0$ and $t$, $N_t$, follows a homogeneous Poisson process with rate $\lambda$. The parameter rate $\theta_t^* := \lambda_t$ will be assumed to follow (\ref{paramito}), with possibly a null-volatility $\sigma_t^{\theta} = 0$ if the homogeneity assumption turns out to be true. Because the econometrician does not have access to the raw data, she can't observe directly the exact time of each event. Instead, she only observes the number of events happening 
on a period (for instance a ten-minute block) $[\tau_{i-1,n}, \tau_{i,n})$, that is $R_{i,n} = N_{\tau_{i,n}}^- - N_{\tau_{i-1,n}}$. If the statistician's assumption of homogeneity is true, the returns are IID. In case of heterogeneity, $N_t$ will be a inhomogeneous Poisson process, and the returns $R_{i,n}$ will most likely be neither identically distributed nor independent.
\end{notIID2}

We need to introduce some notation and definitions. On a given block $i=1, 
\cdots, B_n$ the observed returns will be called $R_{i,n}^{1}, \cdots, R_{i,n}^{h_n}$. Formally, it means that $R_{i,n}^j := R_{(i-1) h_n + j, n}$ for any $j=1, \cdots, h_n$. In analogy with $R_{i,n}^{j}$, we introduce the approximated returns $ \tilde{R}_{i,n}^{1}, \cdots, \tilde{R}_{i,n}^{h_n}$ on the $i$th block. We also introduce the corresponding 
observation times $\tau_{i,n}^{j} := \tau_{(i-1) h_n +j,n}$ for $j=0, \cdots, h_n$. Note that $\tau_{i,n}^0 = 
\tau_{i-1,n}^{h_n}$. Finally, for $j=1, \cdots, h_n$ we define the time increment between the $(j-1)$th return and 
the $j$th return of the $i$th block as $\Delta \tau_{i,n}^j := \tau_{i,n}^j - \tau_{i,n}^{j-1}$. 

\smallskip
We provide a time-varying generalization of the parametric model (\ref{paramret}) as well as a formal expression for the approximated returns. To deal with the former, we assume that in general
\begin{eqnarray}
\label{timevarret} R_{i,n} & := & F_n \big( U_{i,n}, \{ \theta_s^* \}_{\tau_{i-1,n} \leq s \leq \tau_{i,n}} \big).
\end{eqnarray}
The time-varying parameter model in (\ref{timevarret}) is a natural extension of the parametric model (\ref{paramret}) because the returns $R_{i,n}$ can depend on the parameter process path from the previous sampling time $\tau_{i-1,n}$ to the current sampling time $\tau_{i,n}$. As $R_{i,n}$ depend on the parameter path, it seems natural to allow $U_{i,n}$ to be themselves process paths. For example, when the parameter is equal to the volatility process $\theta_t^* := \sigma_t^2$, we will assume that $U_{i,n}$ are equal to the underlying Brownian motion $W_t$ path (see Example \ref{volUn} for more details). Also, as $U_{i,n}$ are random innovation, they should be independent of the parameter process path past, but not on the current parameter path. In the case of volatility, it means that we allow for the leverage effect. A simple particular case of (\ref{timevarret}) is given by
\begin{eqnarray}
\label{timevarrets} R_{i,n} & := & F_n \big( U_{i,n},  \theta_{\tau_{i-1,n}}^* \big),
\end{eqnarray}
i.e. the returns depend on the parameter path only through its initial value. Finally, the approximated returns $\tilde{R}_{i,n}$ follow a mixture of the parametric model (\ref{paramret}) with initial block parameter value. We are now providing a formal definition of our intuition. We assume that 
\begin{eqnarray}
\label{FUT1} R_{i,n}^{j} & := & F_n \big( U_{i,n}^j, \{ \theta_s^* \}_{\tau_{i,n}^{j-1} \leq s \leq \tau_{i,n}^{j}} \big), \\
\label{FUT} \tilde{R}_{i,n}^{j} & := & F_n \big( U_{i,n}^j, \tilde{\Theta}_{i,n} \big) ,
\end{eqnarray}
where the random innovation $U_{i,n}^{j}$ take values on a space $\mathcal{U}_n$ that can be functional\footnote{$\mathcal{U}_n$ is a Borel space, for example the space $\mathcal{C}_1 [0, \Delta \tau_n]$ of $1$-dimensional continuous paths parametrized by time $t \in [0, \tau_n]$.} and that can depend on $n$, $U_{i,n}^j$ are IID for a fixed $n$ but the distribution can depend on $n$, and $F_n(x,y)$ is 
 a non-random function\footnote{\label{fnFn}Let $\mathcal{C}_p (\reels^+)$ be the space of $p$-dimensional continuous paths parametrized by time $t \in \reels^+$, which is a Borel space. Consequently, $\mathcal{U}_n \times \mathcal{C}_p (\reels^+)$ is also a Borel space. We assume that $F_n(x,y)$ is a jointly measurable real-valued function on $\mathcal{U}_n \times \mathcal{C}_p (\reels^+)$. Note that the advised reader will have seen that a priori $\{ \theta_s^* \}_{\tau_{i,n}^{j-1} \leq s \leq \tau_{i,n}^j}$ is defined on $\mathcal{C}_p [0, \tau_n]$ (after translation of the domain by $-\tau_{i,n}^{j-1}$) in (\ref{FUT1}) and $\tilde{\Theta}_{i,n}$ is a vector in (\ref{FUT}), whereas both should be defined on the space $\mathcal{C}_p (\reels^+)$ according to the definition. We match the definitions by extending them as continuous paths on $\reels^+$. Formally, if $\theta_t \in \mathcal{C}_p [0, \tau_n]$, we extend it as $\theta_t := \theta_{\tau_n}$ for all $t > \tau_n$. Similarly, if $\theta \in K$, we extend it as $\theta_t := \theta$ for all $t \geq 0$.}. Note that (\ref{FUT1}) is a mere re-expression of (\ref{timevarret}) using a different notation. For any block $i=1, \cdots, B_n$ and for any observation 
time $j=0, \cdots, h_n$ of the $ith$ block, we define $\mathcal{I}_{i,n}^j$\footnote{Let $(\Omega, \mathcal{F}, P)$ be 
a probability space. Define the \emph{sorted filtration}  $\{ \mathcal{I}_{k,n} \}_{k \geq 0}$ such that for 
any non-negative integer $k$ that we can decompose as $k = (i-1) h_n + j$ where $i \in \{1, \cdots, B_n \}$ and $j \in \{0, \cdots, 
h_n \}$, $\mathcal{I}_{k,n} := \mathcal{I}_{i,n}^j$. 
We assume that $\mathcal{I}_{k,n}$ is a (discrete-time) filtration on 
$(\Omega, \mathcal{F}, P)$. 
In addition, we assume that 
$\{ \theta_s^* \}_{0 \leq s \leq \tau_{i,n}^j}$ and $U_{i,n}^j$ are 
$\mathcal{I}_{i,n}^j$-measurable.} the filtration up to time $\tau_{i,n}^j$. The crucial assumption is that $U_{i,n}^j$ has to be independent 
of the past filtration\footnote{past filtration means up to time $\tau_{i,n}^{j-1}$} (and in particular of $\tilde{\Theta}_{i,n}$). Note that we do not assume any independence between 
the random innovation $U_{i,n}^j$ and the parameter process $\{ \theta_s^* \}_{\tau_{i,n}^{j-1} \leq s \leq \tau_{i,n}^{j}}$. We provide directly the definitions of $F_n$ and $U_{i,n}^j$ in the two toy examples.

\begin{Fn} \label{volUn}
(estimating volatility) In this case, $\mathcal{U}_n$ is defined as the space $\mathcal{C}_1 [0, \Delta \tau_n]$ of continuous paths parametrized by time $t \in [0, \tau_n]$, 
$U_{i,n}^j := \{ \Delta W_{[\tau_{i,n}^{j-1} , s]} 
\}_{\tau_{i,n}^{j-1} \leq s \leq \tau_{i,n}^j}$ are the Brownian motion increment path processes between two consecutive observation times. We assume that $(W_t^{\theta}, W_t)$ is jointly a (possibly non-standard) $2$-dimensional Brownian motion. Thus, the random innovation $U_{i,n}^j$ are indeed independent of the past in view of the Markov property of Brownian motions. We also define  $F_n ( u_t, \theta_t ) := \int_0^{\tau_n} \theta_s^{\frac{1}{2}} du_s$. We thus obtain that the returns are defined as 
$R_{i,n}^j := \int_{\tau_{i,n}^{j-1}}^{\tau_{i,n}^j} \sigma_s dW_s$ and that the approximated returns
 $\tilde{R}_{i,n}^j := \sigma_{\tau_{i,n}^0} \Delta W_{[\tau_{i,n}^{j-1}, \tau_{i,n}^{j}]}$ are the same quantity when holding the volatility constant on the block.
\end{Fn}

\begin{Uijn}
(estimating the rate of a Poisson process) We assume that the rate of the (possibly inhomogeneous) Poisson process is $\alpha_n \lambda_t$, where $\alpha_n$ is a non time-varying and non-random quantity such that $\alpha_n \Delta \tau_{n} := 1$. In this case, we assume that $\mathcal{U}_n$ is the space of increasing paths on $\reels^+$ starting from $0$ which takes values in $\naturels$ and whose jumps are equal to 1. We also assume that for any path in 
$\mathcal{U}_n$, the number of jumps is finite on any compact of $\reels^+$. $U_{i,n}^j$ can be defined as standard Poisson processes 
$\{ N_t^{i,j,n} \}_{t \geq 0}$, independent of each other. We also have $F_n(u_t, \theta_t) := u_{\int_0^{\tau_n} \alpha_n \theta_s du_s}$. Thus, if we let $t_{i,n}^j := \int_{\tau_{i,n}^{j-1}}^{\tau_{i,n}^j} \alpha_n \lambda_s ds$, the returns are the time-changed Poisson processes
\begin{eqnarray}
 \label{ex4R} R_{i,n}^j & = & N_{t_{i,n}^j}^{i,j,n}, \\ 
\label{ex4tildeR} \tilde{R}_{i,n}^j & = & N_{\alpha_n \Delta \tau_{i,n}^j \lambda_{\tau_{i,n}^0}^{i,j,n}}.
 \end{eqnarray} 
\end{Uijn}

\subsection*{Consistency} 
In the following of this section, we will make the block size $h_n$ go to infinity
\begin{eqnarray}
 \label{hninf} h_n \rightarrow \infty.
\end{eqnarray}
Furthermore, we will make the block length $\Delta \Tau_{i,n}$ vanish asymptotically. Because we assume observations are regular in this section, this can be expressed as
\begin{eqnarray}
 \label{hnn-1} h_n n^{-1} \rightarrow 0.
\end{eqnarray}
We can rewrite the consistency of $\widehat{\Theta}_n$ as
\begin{eqnarray}
\label{consistency}
\sum_{i=1}^{B_n} \big( \widehat{\Theta}_{i,n} - \Theta_{i,n} \big) \Delta \Tau_{i,n} \overset{\proba}{\rightarrow} 0.
\end{eqnarray}
where the formal definition of $\widehat{\Theta}_{i,n}$ can be found in (\ref{hatthetain}). In order to show (\ref{consistency}), we can decompose the increments 
$(\widehat{\Theta}_{i,n} - \Theta_{i,n})$ into the part related to misspecified distribution error, the part on 
estimation of approximated returns error and the evolution in the spot parameter error 
\begin{eqnarray}
\label{decompose}
\widehat{\Theta}_{i,n} - \Theta_{i,n} & = & \big( \widehat{\Theta}_{i,n} - \widehat{\tilde{\Theta}}_{i,n} \big) 
+  \big( \widehat{\tilde{\Theta}}_{i,n} - \theta_{\Tau{i-1,n}}^* \big) 
\\ & & + \big( \theta_{\Tau{i-1,n}}^* - \Theta_{i,n} \big) , \nonumber
\end{eqnarray}
where $\widehat{\tilde{\Theta}}_{i,n}$, which is defined formally in (\ref{hattildethetain}), is the parametric estimator used on the underlying non-observed 
approximated returns. It is not a feasible estimator and 
appears in (\ref{decompose}) only to shed light on the way 
we can obtain the consistency of the estimator in the proofs. We first deal with the last error term in (\ref{decompose}), which is due to the non-constancy of the spot parameter $\theta_t^*$. Note that
\begin{eqnarray}
\sum_{i=1}^{B_n} \big( \tilde{\Theta}_{i,n} - \Theta_{i,n} \big) \Delta \Tau_{i,n} 
= \sum_{i=1}^{B_n} \big( \theta_{\Tau_{i-1,n}}^* \Delta \Tau_{i,n} - 
\int_{\Tau_{i-1,n}}^{\Tau_{i,n}} \theta_s^* ds \big)
\end{eqnarray}
and thus we deduce from 
Riemann-approximation\footnote{see i.e. Proposition 4.44 in p.51 of Jacod and Shiryaev (2003)} that
\begin{eqnarray}
\label{tilde} \sum_{i=1}^{B_n} \big( \tilde{\Theta}_{i,n} - \Theta_{i,n} \big) \Delta \Tau_{i,n} \overset{\proba}{\rightarrow} 0.
\end{eqnarray}
To deal with the other terms in (\ref{decompose}), we assume that for any positive integer $k$, the practitioner has at hand an estimator $\hat{\theta}_{k,n} := \hat{\theta}_{k,n} 
(r_{1,n} ;$ $\cdots ;$ $r_{k,n} )$,
which depends on the input of returns $\{ r_{1,n} ; \cdots ; r_{k,n} \}$. On each block $i=1, \cdots, B_n$ we estimate the local parameter as
\begin{eqnarray}
 \label{hatthetain} \widehat{\Theta}_{i,n} := \hat{\theta}_{h_n,n} \big( R_{i,n}^1 ; \cdots ; 
 R_{i,n}^{h_n} \big) .
\end{eqnarray}
The non-feasible estimator $\widehat{\tilde{\Theta}}_{i,n}$ is defined as
the same parametric estimator with approximated returns as input instead of observed returns
\begin{eqnarray}
 \label{hattildethetain} \widehat{\tilde{\Theta}}_{i,n} := \hat{\theta}_{h_n,n} \big( \tilde{R}_{i,n}^1 ; \cdots ; 
 \tilde{R}_{i,n}^{h_n} \big) .
\end{eqnarray}
Note that (\ref{hattildethetain}) is infeasible because the approximated returns $\tilde{R}_{i,n}^j$ are non-observable quantities. 

\begin{estim1}
 (estimating volatility) The estimator 
is the scaled usual RV, i.e. 
$\hat{\theta}_{k,n} (r_{1,n} ;$ $\cdots ; r_{k,n} )$ $:= T^{-1} k^{-1} n \sum_{j=1}^{k} r_{j,n}^2$. Note that 
$\hat{\theta}_{k,n}$ can also be seen as the MLE (see the discussion pp. 112-115 in Mykland and Zhang (2012)).
\end{estim1}

\begin{estim2}
 (estimating the rate of a Poisson process) The estimator to be 
used is the return mean $\hat{\theta}_{k,n} (r_{1,n} ; \cdots ; r_{k,n} ) := k^{-1} \sum_{j=1}^{k} r_{j,n}$.
\end{estim2}
In order to tackle the second term in (\ref{decompose}), we make 
the assumption that the parametric estimator is $\mathbf{L}^1$-convergent, locally uniformly in the model parameter $\theta$ if 
we actually observe returns coming from the parametric model. This can be expressed in the following condition.

\begin{conditionC1}
 Let the innovation of a block $(V_{1,n}, \cdots, V_{h_n,n})$ be IID with distribution $U_n$. For any $M > 0$,
 \begin{eqnarray*}
\label{C1rk} \underset{\theta \in K_M}{\sup} \text{ } \esp \left[ \big| \hat{\theta}_{h_n,n} 
( F_n( V_{1,n}, \theta)  
; \cdots ; F_n(V_{h_n,n}, \theta)  ) - \theta \big| \right] \rightarrow 0.
\end{eqnarray*}
\end{conditionC1}
\begin{rkC1} \label{rkC1}
(practicability) Under Condition (C), results on regular conditional 
distributions\footnote{see for instance Leo Breiman (1992), see Section \ref{proofs} for more details.} give us that 
the error made on the estimation of the underlying non-observed returns tends to 0, i.e.
\begin{eqnarray}
\label{tildehat} \sum_{i=1}^{B_n} \big( \widehat{\tilde{\Theta}}_{i,n} - \tilde{\Theta}_{i,n} \big) \Delta \Tau_{i,n} \overset{\proba}{\rightarrow} 0.
\end{eqnarray}
This proof technique is the main idea of the paper. Regular conditional distributions are used to deduce results on the time-varying parameter model using uniform results in the parametric model.  
\end{rkC1}
\begin{conditionC1rk}
 (consistency) Note that $\mathbf{L}^1$-convergence is slightly stronger than the simple consistency of the parametric estimator. Nonetheless, in most applications, we will have both.
\end{conditionC1rk}

We can now summarize the consistency result in this very simple case where observations occur at equidistant time intervals and returns are IID under the parametric model.
 Under Condition (C) and assuming that \begin{eqnarray}
\label{discrephat} \sum_{i=1}^{B_n} (\widehat{\Theta}_{i,n} - \widehat{\tilde{\Theta}}_{i,n}) \Delta \Tau_{i,n} \overset{\proba}{\rightarrow} 0,
\end{eqnarray} 
we have the consistency of (\ref{newestimator}), i.e. 
 \begin{eqnarray}
 \label{consistencyappendix} \widehat{\Theta}_{n} \overset{\proba}{\rightarrow} \Theta.
 \end{eqnarray}
We obtain the consistency in the couple of toy examples\footnote{see Section \ref{proofs} for proofs}.

\begin{notIIDexamples} (LPE equal to the parametric estimator) The reader will have noticed that in the couple of examples, the LPE is equal to the parametric estimator. This is because in those very basic examples, the parametric estimator is linear, i.e. for any positive integer $k$ and 
$l = 1, \cdots, k-1$
\begin{eqnarray*}
 \hat{\theta}_{k,n} (r_{1,n} ; \cdots ; r_{k,n} ) = \frac{l}{k} \hat{\theta}_{l,n} (r_{1,n} ; \cdots ; r_{l,n} ) 
+ \frac{k-l}{k} \hat{\theta}_{k-l,n} (r_{l+1,n} ; \cdots ; r_{k,n} )
\end{eqnarray*}
In more general examples, this equation will break, and we will obtain two distinct estimators. 
\end{notIIDexamples}

\pagebreak
\section{Proofs}
\label{proofs}
\subsection{Preliminaries}
In view of our assumptions on $\theta_t^*$, we can follow standard 
localisation arguments (see, e.g., pp. $160-161$ of Mykland and Zhang (2012)) and assume without loss of generality 
that $K$ is a compact space. In case $\theta_t^*$ is an It\^{o} semimartingale satisfying Condition (P1), we can also assume without loss of generality that there exists $0 \leq \sigma^+$ such that for any eigen value $\lambda_t$ of $\sigma_t^{\theta}$, we have $0 \leq \lambda_t \leq \sigma^+$ and that there exists $0 \leq a^+$ such that $\mid a_t^{\theta} \mid \leq a^+$.

\smallskip
Finally, we fix some notation. In the following of this paper, we will be using $C$ for any constant $C > 0$, where the value can change from one line to the next. 

\smallskip
We start with the proofs related to the consistency in the simple model introduced in Section \ref{cons}. This provides an overview of the proof techniques, although the techniques will be more intricate when proving Theorem \ref{clt} (Central limit theorem), which includes non-regular observations.

\subsection{Proof of Condition (C) $\Rightarrow $(\ref{tildehat})} \label{consistproof}
It is sufficient to show that Condition (C) implies that 
\begin{eqnarray}
\label{proofAC2} \underset{i \geq 0}{\sup} \text{ } \esp \left[ \big| \widehat{\tilde{\Theta}}_{i,n} - 
\theta_{\Tau{i-1,n}}^* \big| \right] = o_p (1).
\end{eqnarray}
By (\ref{FUT}) and (\ref{hattildethetain}), we can build $g_n$ such that we can write 
$$\big| \widehat{\tilde{\Theta}}_{i,n} - \theta_{\Tau{i-1,n}}^* \big| = g_n (U_{i,n}^1, \cdots, U_{i,n}^{h_n}, 
\theta_{\Tau{i-1,n}}^*),$$
where $g_n$ is a jointly measurable real-valued function such that 
$$\esp \big| 
g_n (U_{i,n}^1, \cdots, U_{i,n}^{h_n}, \theta_{\Tau{i-1,n}}^* \big| < \infty.$$
We have that 
\begin{eqnarray*}
 \esp \left[ g_n (U_{i,n}^1, \cdots, U_{i,n}^{h_n}, \theta_{\Tau{i-1,n}}^*) \right] & = & 
 \esp \left[ \esp \left[ g_n (U_{i,n}^1, \cdots, U_{i,n}^{h_n}, \theta_{\Tau{i-1,n}}^*) \big| \theta_{\Tau{i-1,n}}^*\right] \right]\\
 & = & \esp \left[ \int g_n (u, \theta_{\Tau{i-1,n}}^*) \mu_{\omega} (du) \right]
\end{eqnarray*}
where $\mu_{\omega} (du)$ is a regular conditional distribution for $(U_{i,n}^1, \cdots, U_{i,n}^{h_n})$ given 
$\tilde{\Theta}_{i,n}$ (see, 
e.g., Breiman (1992)). From Condition (C), we obtain (\ref{proofAC2}).

\subsection{Proof of the consistency in Example \ref{estimatingvolatility}} \label{proofconsistex}
Let's show Condition (C) first. For any $M > 0$, the quantity
$$\Big| \hat{\theta}_{h_n,n} \big( F_n ( V_{1,n} , \theta ) ; \cdots ; F_n ( V_{h_n,n} , \theta ) \big) - \theta \Big|$$
can be shown to go to 0 in probability as a straightforward consequence of Theorem $I.4.47$ of p.52 in 
Jacod and Shiryaev (2003).

\smallskip
To show the condition (\ref{discrephat}), it is sufficient to show that the following quantity 
\begin{eqnarray}
\label{consistency13} n h_n^{-1} \esp_{\Tau_{i-1,n}} \left[ \Big| \big( \theta_{\Tau_{i-1,n}}^* 
\Delta W_{[\Tau_{i-1,n} ; \Tau_{i,n}]} \big)^2 - \big( \int_{\Tau_{i-1,n}}^{\Tau_{i,n}} 
\theta_s^* d W_s \big)^2 \Big| \right]
\end{eqnarray}
goes to $0$ uniformly in i. To prove this, we can use the formula $(a^2 - b^2) = (a+b)(a-b)$, together with conditional Burkholder-Davis-Gundy inequality 
(BDG, see inequality (2.1.32) of p. 39 in Jacod and Protter (2011)).

\subsection{Proof of Consistency in Example \ref{estimatingratepoisson}}
Condition (C) can be shown easily. Similarly, the condition (\ref{discrephat}) is a direct consequence of the definition in (\ref{ex4R}), (\ref{ex4tildeR}) together with (\ref{hnn-1}).

\subsection{Proof of Theorem \ref{clt} (Central limit theorem with non regular observation times)}
We prove directly the central limit theorem in this general case. As a by-product, this implies the case with regular observations, i.e. Theorem \ref{cltreg}.
We can decompose $n^{\frac{1}{2}} \sum_{i=1}^{B_n} \big( \widehat{\Theta}_{i,n} - \Theta_{i,n} \big) \Delta \Tau_{i,n}$ as 
\begin{eqnarray}
\label{decomposeCLT0} 
I + II + III + IV,
\end{eqnarray}
with 
\begin{eqnarray*}
I & = &n^{\frac{1}{2}} \sum_{i=1}^{B_n} \big(\widehat{\Theta}_{i,n} \Delta \Tau_{i,n} - 
\hat{\tilde{\Theta}}_{i,n}^{\mathbf{P}} \Delta \widetilde{\Tau}_{i,n}^{\mathbf{P}} \big),\\ II & = & n^{\frac{1}{2}} \sum_{i=1}^{B_n} \big( \hat{\tilde{\Theta}}_{i,n}^{\mathbf{P}} - \theta_{\Tau_{i-1,n}}^* \big) \Delta \widetilde{\Tau}_{i,n}^{\mathbf{P}},\\ III & = &n^{\frac{1}{2}} \sum_{i=1}^{B_n} \theta_{\Tau_{i-1,n}}^* \big( \Delta \widetilde{\Tau}_{i,n}^{\mathbf{P}} - \Delta \Tau_{i,n} \big),\\ IV & = &n^{\frac{1}{2}} \sum_{i=1}^{B_n} \big( \theta_{\Tau_{i-1,n}}^* - \Theta_{i,n} \big) \Delta \Tau_{i,n}.
\end{eqnarray*}
It is clear that $I \overset{\proba}{\rightarrow} 0$ by (\ref{nonregeq}) and $III \overset{\proba}{\rightarrow} 0$ by (\ref{nonregeq2}) along with Lemma 2.2.10 (p. 55) in Jacod and Protter (2011) and the fact that $\theta_t^*$ takes values in a compact set. We prove in what follows that $IV \overset{\proba}{\rightarrow} 0$ and that $II \rightarrow \widetilde{Z}$, where $\widetilde{Z}$ follows the definition of Theorem \ref{clt}.
\subsubsection*{We show $IV \overset{\proba}{\rightarrow} 0$}
We consider first the case where $\theta_t^*$ satisfies Condition (P2). We introduce
\begin{eqnarray}
\label{proofACLT1} e_{i,n} & := &  n^{\frac{1}{2}} \big( \theta_{\Tau_{i-1,n}}^* - 
\Theta_{i,n} \big) \Delta \Tau_{i,n}.
\end{eqnarray}
It is sufficient to show that $\sum_{i=1}^{B_n} \mid e_{i,n} \mid \overset{\proba}{\rightarrow} 0$, and by virtue of Lemma 2.2.10 (p. 55) in Jacod and Protter (2011) that $\sum_{i=1}^{B_n} \esp_{\Tau_{i-1,n}} \big[ \mid e_{i,n} \mid \big] \overset{\proba}{\rightarrow} 0$. We compute
\begin{eqnarray*}
\sum_{i=1}^{B_n} \esp_{\Tau_{i-1,n}} \big[\mid e_{i,n} \mid \big] 
 & = & n^{\frac{1}{2}} \sum_{i=1}^{B_n} \esp_{\Tau_{i-1,n}} 
 \Big[ \big| \int_{\Tau_{i-1,n}}^{\Tau_{i,n}} ( \theta_u^* - 
 \theta_{\Tau_{i-1,n}}^*) du \big| \Big] \\
 & \leq & C n^{\frac{1}{2}} \sum_{i=1}^{B_n} \underbrace{\big( \esp_{\Tau_{i-1,n}} \big[ ( \Delta \Tau_{i,n} )^2 
 \big] \big)^{\frac{1}{2}} }_{O_p (h_n n^{-1})} \\ & & \underbrace{\Big( \esp_{\Tau_{i-1,n}} \big[ \underset{\Tau_{i-1,n} \leq s \leq \Tau_{i,n}}{\sup} 
 \big| \theta_s^* - \theta_{\Tau_{i-1,n}}^* \big|^2  
 \big] \Big)^{\frac{1}{2}} }_{o_p (n^{-\frac{1}{2}})} \\
 & = & o_p (1),
\end{eqnarray*}
where we used onditional Cauchy-Schwarz to obtain the inequality, Condition (T) along with Condition (P2) to obtain the last equality. We deduce that $IV \overset{\proba}{\rightarrow} 0$ in this case too.

\smallskip
We now consider the case where $\theta_t^*$ satisfies Condition (P1) and (\ref{E21}) holds. We start by decomposing $e_{i,n}$ into its bias and its martingale part. We have
$$e_{i,n} = \underbrace{n^{\frac{1}{2}} \int_{\Tau_{i-1,n}}^{\Tau_{i,n}} \int_{\Tau_{i-1,n}}^{s} a_u^\theta du ds}_{e_{i,n}^{(b)}} + \underbrace{n^{\frac{1}{2}} \int_{\Tau_{i-1,n}}^{\Tau_{i,n}} \int_{\Tau_{i-1,n}}^{s} \sigma_u^\theta dW_u ds}_{e_{i,n}^{(m)}}.$$
We will show in what follows that $\sum_{i=1}^{B_n} e_{i,n}^{(b)} = o_\proba (1)$ and $\sum_{i=1}^{B_n} e_{i,n}^{(m)} = o_\proba (1)$. We start with the first assertion. As for the previous case, it is sufficient to show that $\sum_{i=1}^{B_n} \esp_{\Tau_{i-1,n}} \big[ \mid e_{i,n}^{(b)} \mid \big] \overset{\proba}{\rightarrow} 0$. As $a_t^{\theta}$ is bounded, we can bound the expression via
\begin{eqnarray*}
\sum_{i=1}^{B_n} \esp_{\Tau_{i-1,n}} \big[\mid e_{i,n}^{(b)} \mid \big] 
 & \leq & C n^{\frac{1}{2}} \sum_{i=1}^{B_n} \esp_{\Tau_{i-1,n}} 
 \big[ (\Delta \Tau_{i,n})^2 \big].
\end{eqnarray*}
Then, using Condition (T) along with (\ref{E21}), we conclude that this is $o_\proba (1)$.

\smallskip
We show now that $\sum_{i=1}^{B_n} e_{i,n}^{(m)} = o_\proba (1)$. As it is a martingale, it is sufficient to show that $\sum_{i=1}^{B_n} \esp_{\Tau_{i-1,n}} \big[ \mid e_{i,n}^{(m)} \mid^2 \big] \overset{\proba}{\rightarrow} 0$. We compute 
\begin{eqnarray*}
\sum_{i=1}^{B_n} \esp_{\Tau_{i-1,n}} \big[\mid e_{i,n}^{(m)} \mid^2 \big] 
 & = & n \sum_{i=1}^{B_n} \esp_{\Tau_{i-1,n}} 
 \Big[ \Big| \int_{\Tau_{i-1,n}}^{\Tau_{i,n}} \int_{\Tau_{i-1,n}}^{s} \sigma_u^\theta dW_u ds \Big|^2 \Big] \\
 & \leq & C n \sum_{i=1}^{B_n} \big( \esp_{\Tau_{i-1,n}} \big[ ( \Delta \Tau_{i,n} )^3 
 \big] \big)^{\frac{2}{3}}  \\ & & \Big( \esp_{\Tau_{i-1,n}} \Big[ \underset{\Tau_{i-1,n} \leq s \leq \Tau_{i,n}}{\sup} 
 \Big| \int_{\Tau_{i-1,n}}^{s} \sigma_u^\theta dW_u \Big|^6  
 \Big] \Big)^{\frac{1}{3}}, \\
  & \leq & C n \sum_{i=1}^{B_n}  \esp_{\Tau_{i-1,n}} \big[ ( \Delta \Tau_{i,n} )^3 
 \big] \\
 & = & o_p (1),
\end{eqnarray*}
where we used conditional H\"{o}lder's inequality with $p=3/2$ and $q=3$ in the first inequality, BDG with $p=3$ in the second inequality, Condition (T) along with (\ref{E21}) in the last equality.

\subsubsection*{We show $II \rightarrow \widetilde{Z}$}
We aim to use Theorem 2-2 (p. 242) in Jacod (1997). Conditions are further specified in Theorem 3-2 (p. 244) in the case when observations are regular. Following the proof of Theorem 3-2, we can actually show that such conditions hold in the more general case when observations are not regular, choosing the filtration $\mathcal{J}_{\Tau_{i,n}}$. It is crucial to note that we are not working with the filtration $\mathcal{J}_{\tau_{i,n}}$. 

\smallskip
Consequently, our goal is to show the conditions (3.10)-(3.14) from Theorem 3-2 (p. 244) in Jacod (1997). Note that (3.12) and (3.14) are respectively implied by (\ref{condnotreg1}) and (\ref{condnotreg2}). The bias condition (3.10) is satisfied as an application of (\ref{E611}) along with regular conditional distribution. 

\smallskip
In this step, we prove that (3.11) is satisfied. We introduce $A_{i,n} := n^{\frac{1}{2}} \big( \hat{\tilde{\Theta}}_{i,n}^{\mathbf{P}} - \theta_{\Tau_{i-1,n}}^* \big) \Delta \widetilde{\Tau}_{i,n}^{\mathbf{P}}$ and 
$$C_{i,n} := \esp_{\Tau_{i-1,n}} \big[ A_{i,n} A_{i,n}^T \big] - \esp_{\Tau_{i-1,n}} \big[ A_{i,n}  \big] \esp_{\Tau_{i-1,n}} \big[ A_{i,n}^T \big].$$ 
The condition (3.11) can be expressed as  
\begin{eqnarray}
 \label{CLT32} \sum_{i=1}^{B_n}  C_{i,n}  
 \overset{\proba}{\rightarrow} T \int_0^T V_{\theta_s^*} ds.
\end{eqnarray}
By regular conditional distribution, (\ref{E611}) and (\ref{E321}), we have that 
\begin{eqnarray*}
 \sum_{i=1}^{B_n} C_{i,n}  & = & 
 T \sum_{i=1}^{B_n} \esp_{\Tau_{i-1,n}} \big[ V_{\theta_{\Tau_{i-1,n}}^*} \Delta \widetilde{\Tau}_{i,n}^{\mathbf{P}} \big] 
 + o_p (1).
 \end{eqnarray*}
In view of (\ref{nonregeq2}), the conditional Cauchy-Schwarz inequality and the boundedness of $V_{\theta}$, we get
$$\sum_{i=1}^{B_n} \esp_{\Tau_{i-1,n}} \left[ V_{\theta_{\Tau_{i-1,n}}^*} \Delta \widetilde{\Tau}_{i,n}^{\mathbf{P}} \right] = \sum_{i=1}^{B_n} \esp_{\Tau_{i-1,n}} \left[ V_{\theta_{\Tau_{i-1,n}}^*} \Delta \Tau_{i,n} \right] + o_p (1).$$
Using Lemma 2.2.11 of Jacod and Protter (2011) together with conditional Cauchy-Schwarz inequality, (\ref{obs1}) and the boundedness of $V_{\theta}$, we obtain
$$T \sum_{i=1}^{B_n} \esp_{\Tau_{i-1,n}} \left[ V_{\theta_{\Tau_{i-1,n}}^*} \Delta \Tau_{i,n} \right] = T \sum_{i=1}^{B_n} V_{\theta_{\Tau_{i-1,n}}^*} \Delta \Tau_{i,n} + o_p (1).$$
We can apply now Proposition I.4.44 (p. 51) in Jacod and Shiryaev (2003) and we get
\begin{eqnarray*}
T \sum_{i=1}^{B_n} V_{\theta_{\Tau_{i-1,n}}^*} \Delta \Tau_{i,n} 
& \overset{\proba}{\rightarrow} & T \int_0^T V_{\theta_s^*} ds.
\end{eqnarray*}

\smallskip
In this final step, we prove that the Lindeberg condition (3.13) is satisfied. We will show in this step that for all $\epsilon > 0$,
\begin{eqnarray}
 \label{CLT310} \sum_{i=1}^{B_n} \esp_{\Tau_{i-1,n}} \left[ \mid A_{i,n} \mid^2 \mathbf{1}_{\{ \mid A_{i,n} \mid > \epsilon \}} \right] 
 \overset{\proba}{\rightarrow} 0.
\end{eqnarray}
Actually, (\ref{CLT310}) can be shown using regular conditional distribution along with (\ref{E331}).

\subsection{Proof of Theorem \ref{thQMLE} (QMLE)}
We want to show that the conditions of Theorem \ref{cltreg} are satisfied. We start with the case $\alpha > \frac{1}{2}$. The key result is Theorem 6 in Xiu (2010, p. 241). We choose $\mathbf{P} = (0,0)$.

\smallskip
We show first Condition (E). We can see easily from the key result that if we choose $V_{\theta_t^*} = 6 \sigma_t^2$, then (\ref{E321}) is satisfied. 

\smallskip
We can verify the Lindeberg condition (\ref{E331}) using conditional Cauchy-Schwarz inequality and the fact that the fourth moment of $$h_n^{\frac{1}{2}} \big( \hat{\tilde{\Theta}}_{i,n}^{\mathbf{P},\theta} - \theta \big)$$ 
is bounded. 

\smallskip
As for the bias condition (\ref{E611}), we can see that as the noise shrinks faster than the order of the returns to $0$, then the bias tends to the sum of the diagonal elements of $W_1$ defined in (23) in Xiu (2010, p. 241) minus unity. This equals $0$ and thus (\ref{E611}) is satisfied. 

\smallskip
The condition (\ref{condE+}) is satisfied combining the fact that the noise is independent from $X_t$, the  aforementioned theorem with the rationale in Section \ref{proofconsistex}.

\smallskip
We now show that (\ref{Gt}) and (\ref{Nt}) are satisfied. Actually, we can show trivially that (\ref{Gt}) holds for the reference continuous martingale $M_t = 0$. We recall that we are "only" showing stable convergence with respect to $\mathcal{F}_t^X$, and we show now the condition related to stable convergence (\ref{Nt}). Actually, we can assume that $N_t = X_t$, this will imply that the result holds for any $N \in \mathcal{M}_b(M^{\perp})$. From Theorem 6 in Xiu (2010), we have that 
$$\hat{\tilde{\Theta}}_{i,n}^{\mathbf{P}} = \sum_{k=(i-1)h_n}^{i h_n-1} \sum_{l=k+1}^{ih_n-1} \omega_{k,l,n} (Z_{\tau_{k+1,n},n} - Z_{\tau_{k,n},n}) (Z_{\tau_{l+1,n},n} - Z_{\tau_{l,n},n}). $$
We can develop $(Z_{\tau_{k+1,n},n} - Z_{\tau_{k,n},n}) (Z_{\tau_{l+1,n},n} - Z_{\tau_{l,n},n}) = I + II + III + IV$, where $I=(X_{\tau_{k+1,n}} - X_{\tau_{k,n}}) (X_{\tau_{l+1,n}} - X_{\tau_{l,n}})$, $II= (X_{\tau_{k+1,n}} - X_{\tau_{k,n}}) (\epsilon_{l+1,n} - \epsilon_{l,n})$, 
$III =(\epsilon_{k+1,n} - \epsilon_{k,n}) (X_{\tau_{l+1,n}} - X_{\tau_{l,n}})$ and $IV= (\epsilon_{k+1,n} - \epsilon_{k,n}) (\epsilon_{l+1,n} - \epsilon_{l,n})$. Because the noise is independent from $X_t$, it is clear that $\esp_{\Tau_{i-1,n}} \big[II*( X_{\Tau_{i,n}} - X_{\Tau_{i-1,n}}) \big] = 0$, $\esp_{\Tau_{i-1,n}} \big[ III*( X_{\Tau_{i,n}} - X_{\Tau_{i-1,n}}) \big] = 0$ and $ \esp_{\Tau_{i-1,n}} \big[ IV*( X_{\Tau_{i,n}} - X_{\Tau_{i-1,n}}) \big]= 0$. As for $I$, we can express
$$I*( X_{\Tau_{i,n}} - X_{\Tau_{i-1,n}}) = I*\sum_{k=(i-1)h_n}^{i h_n-1} ( X_{\tau_{k+1,n}} - X_{\tau_{k,n}})$$
and from this expression straightforward computation leads to
$$\esp_{\Tau_{i-1,n}} \Big[\big( \hat{\tilde{\Theta}}_{i,n}^{\mathbf{P}} - \tilde{\Theta}_{i,n} \big) \big( X_{\Tau_{i,n}} - X_{\Tau_{i-1,n}} \big) \Big] = 0.$$

\smallskip
We consider now the case $\alpha = 1/2$, i.e. when both the noise variance and the returns are of the same rate. In that case, we need to use the bias-corrected estimator $\widehat{\Theta}_n^{(BC)}$ so that we can verify the conditions of Theorem \ref{cltreg}. The key result here is Proposition 1 (p. 369) along with its proof (p. 391-393) in A\"{i}t-Sahalia et al. (2005). 

\smallskip
The bias condition (\ref{E611}) is satisfied on the account that we have reduced the bias of the estimator. Actually, the de-bias of the estimator doesn't affect the rest of the proof. Moreover, the increment condition (\ref{E321}) and the Lindeberg condition (\ref{E331}) are satisfied using similar techniques of proof. Finally, the conditions (\ref{Gt}), (\ref{Nt}) and (\ref{condE+}) are satisfied using the same line of reasoning as in the previous case.

\subsection{Proof of Theorem \ref{thQMLEpowers} (powers of volatility)}
We aim to show that we can verify the conditions of Theorem \ref{cltreg}. The idea is to use a Taylor expansion as in the delta method. Then the conditions will be satisfied partly following the proof of Theorem \ref{thQMLE}. More specifically, we will do the proof for (\ref{E6110}) and (\ref{E3210}) of Condition (E), but will not explicit the proof of (\ref{E3310}), (\ref{Gt}), (\ref{Nt}), (\ref{condE+}) which can be proven using the same ideas. We use the following notation:
\begin{eqnarray}
\hat{\tilde{\Theta}}_{i,n}^{\mathbf{P},\theta} := g(\hat{\tilde{\sigma}}_{i,n}^{2,\mathbf{P},\sigma}) - B_{i,n},
\end{eqnarray}
where $B_{i,n}$ can correspond to either one of the two bias-correction expressions found in (\ref{1608}) and (\ref{1609}). We have that 
\begin{eqnarray}
\label{081600}\hat{\tilde{\Theta}}_{i,n}^{\mathbf{P},\theta} - \theta := g(\hat{\tilde{\sigma}}_{i,n}^{2,\mathbf{P},\sigma}) - B_{i,n} - g(\sigma^2),
\end{eqnarray}
for some $\sigma^2$. Using a Taylor expansion, we obtain that:
\begin{eqnarray}
\nonumber g(\hat{\tilde{\sigma}}_{i,n}^{2,\mathbf{P},\sigma}) - g(\sigma^2) & = & (\hat{\tilde{\sigma}}_{i,n}^{2,\mathbf{P},\sigma} - \sigma^2)g'(\sigma^2) + \frac{1}{2}(\hat{\tilde{\sigma}}_{i,n}^{2,\mathbf{P},\sigma} - \sigma^2)^2g''(\sigma^2) \\ \label{taylor}& & + \frac{1}{6}(\hat{\tilde{\sigma}}_{i,n}^{2,\mathbf{P},\sigma} - \sigma^2)^3g^{(3)}(\eta),
\end{eqnarray}
where $\eta$ is between $\sigma^2$ and $\hat{\tilde{\sigma}}_{i,n}^{2,\mathbf{P},\sigma}$. Combining (\ref{081600}) and (\ref{taylor}) and several assumptions (including the conditions on $g$), we obtain:
 \begin{eqnarray}
 \var \big[ h_n^{\frac{1}{2}} \big( \hat{\tilde{\Theta}}_{i,n}^{\mathbf{P},\theta} - \theta \big) \big]  & = & (g'(\sigma^2))^2 \var \big[ h_n^{\frac{1}{2}} (\hat{\tilde{\sigma}}_{i,n}^{2,\mathbf{P},\sigma} - \sigma^2) \big] + o (1).
  \end{eqnarray}
 From here, one can conclude (\ref{E3210}) using the proof of Theorem \ref{thQMLE}.
 
 \smallskip
 As for the bias condition (\ref{E6110}), combining (\ref{081600}) and (\ref{taylor}) and several assumptions we deduce that:
 \begin{eqnarray}
\nonumber \esp \big[ \big( \hat{\tilde{\Theta}}_{i,n}^{\mathbf{P},\theta} - \theta \big) \big]  & = & \esp \big[(\hat{\tilde{\sigma}}_{i,n}^{2,\mathbf{P},\sigma} - \sigma^2)g'(\sigma^2) + \frac{1}{2}(\hat{\tilde{\sigma}}_{i,n}^{2,\mathbf{P},\sigma} - \sigma^2)^2g''(\sigma^2) - B_{i,n} \big] \\  \label{001} & & + o (n^{-\frac{1}{2}}).
  \end{eqnarray}
We can show that 
\begin{eqnarray}
\label{002} \esp \big[(\hat{\tilde{\sigma}}_{i,n}^{2,\mathbf{P},\sigma} - \sigma^2)g'(\sigma^2) \big]= o (n^{-\frac{1}{2}})
\end{eqnarray}
as in the proof of Theorem \ref{thQMLE}. We can also show that 
\begin{eqnarray}
\label{003}
\esp \big[\frac{1}{2}(\hat{\tilde{\sigma}}_{i,n}^{2,\mathbf{P},\sigma} - \sigma^2)^2g''(\sigma^2) - B_{i,n} \big] = o (n^{-\frac{1}{2}})
\end{eqnarray}
following the same line of reasoning as that of the case $v=4$ in the proof of Lemma 4.4 in Jacod and Rosenbaum (2013, p. 1480). In view of (\ref{001}), (\ref{002}) and (\ref{003}), we can show the bias condition (\ref{E6110}).
  
\subsection{Proof of Theorem \ref{thEQMLE} (E-(LPE of QMLE))}
The strategy of the proof consists in showing that the estimation error in $\nu$ does not affect asymptotically the behavior of the QMLE so that we can apply directly Theorem \ref{thQMLE}. To do that, the key results will be Theorem 3(i) (p. 37) in Li et al. (2016) and Theorem 6 (p. 241) in Xiu (2010).

\smallskip
We recall that $\widehat{X}_{\tau_{i,n}} = Z_{\tau_{i,n},n} - g(I_{i,n}, \hat{\nu})$ and we define the n-dimensional vector $\widehat{Y}_n = (\widehat{X}_{\tau_{1,n}} - \widehat{X}_{0}, \cdots, \widehat{X}_{T} - \widehat{X}_{\tau_{n-1,n}})$. We also define $Y_n =  ((X_{\tau_{1,n}} - X_{0}) + (\tilde{\epsilon}_{1,n} - \tilde{\epsilon}_{0,n}), \cdots, (X_{T} - X_{\tau_{n-1,n}})  + (\tilde{\epsilon}_{n,n} - \tilde{\epsilon}_{n-1,n}))$ and $\delta_n = (g(I_{1,n}, \hat{\nu}) - g(I_{0,n}, \hat{\nu})) - (g(I_{1,n}, \nu) - g(I_{0,n}, \nu)), \cdots, g(I_{n,n}, \hat{\nu}) - g(I_{n-1,n}, \hat{\nu})) - (g(I_{n,n}, \nu) - g(I_{n-1,n}, \nu)))$. It is clear that 
\begin{eqnarray}
\label{Ydelta}
\widehat{Y}_n = Y_n + \delta_n. 
\end{eqnarray}
Finally, we recall that $\widehat{\Theta}_n$ is the LPE of QMLE on $\widehat{Y}_n$ and we define $\widetilde{\Theta}_n$ as the LPE of QMLE on $Y_n$.

\smallskip
Consider the case $\alpha < 1/2$ (the case $\alpha = 1/2$ is done following the same line of reasoning). The goal is to show that stably in distribution 
\begin{eqnarray}
\label{goal1}
n^{1/2}\Big(\widehat{\Theta}_n - \Theta \Big) \rightarrow \Big(6 T^{-1} \int_0^T \sigma_s^4 ds \Big)^{\frac{1}{2}} \mathcal{N} (0,1).
\end{eqnarray}
We decompose the left hand-side term in (\ref{goal1}) as 
$$n^{1/2}(\widehat{\Theta}_n - \Theta ) = \underbrace{n^{1/2}(\widetilde{\Theta}_n - \Theta )}_{A_n} + \underbrace{n^{1/2}(\widehat{\Theta}_n - \widetilde{\Theta}_{n})}_{B_n}.$$
 On the account of Theorem \ref{thQMLE}, we have that $A_n \rightarrow (6 T^{-1} \int_0^T \sigma_s^4 ds )^{\frac{1}{2}} \mathcal{N} (0,1).$ Thus, if we can show that $B_n \overset{\proba}{\rightarrow} 0$, then this implies (\ref{goal1}).

\smallskip
We show now that $B_n \overset{\proba}{\rightarrow} 0$. We define $\mathcal{M}_{n}$ the set of real $n \times n$ matrices. In view of Theorem 6 (p. 241) in Xiu (2010), there exists a function 
\begin{eqnarray*}
M  :  K & \rightarrow & \mathcal{M}_{n} \times \mathcal{M}_{n} \\
\theta & \mapsto & (M^{(1)} (\theta), M^{(2)} (\theta))
\end{eqnarray*}
such that $\widehat{\Theta}_{i,n}=\widehat{Y}_n'M (\widehat{\Theta}_{i,n}) \widehat{Y}_n$ and $\tilde{\Theta}_{i,n}=Y_n'M (\tilde{\Theta}_{i,n}) Y_n$, where we define for any $\theta \in K$ and any n dimensional vector $Y$:
$$Y' M(\theta) Y = (Y' M^{(1)}(\theta) Y, Y' M^{(2)}(\theta) Y).$$ 
We have that
\begin{eqnarray*}
B_n & = & n^{1/2}(\widehat{\Theta}_n - \tilde{\Theta}_{i,n}),\\
& = & n^{1/2}(\widehat{Y}_n'M (\widehat{\Theta}_{i,n})\widehat{Y}_n - Y_n'M (\tilde{\Theta}_{i,n}) Y_n),\\
& = & n^{1/2}((Y_n  + \delta_n)'M (\widehat{\Theta}_{i,n}) (Y_n  + \delta_n) - Y_n'M (\tilde{\Theta}_{i,n}) Y_n),\\
& = & n^{1/2}(Y_n'(M (\widehat{\Theta}_{i,n}) - M (\tilde{\Theta}_{i,n})) Y_n  \\
& &+ (\delta_n'M (\widehat{\Theta}_{i,n}) Y_n + Y_n'M (\widehat{\Theta}_{i,n}) \delta_n +  \delta_n'M (\widehat{\Theta}_{i,n}) \delta_n)),\\
& = & n^{1/2}Y_n'(M (\widehat{\Theta}_{i,n}) - M (\tilde{\Theta}_{i,n})) Y_n  + o_p(1),\\
& = & o_p(1).\\
\end{eqnarray*}
where we used (\ref{Ydelta}) in the third equality, Assumption A along with Theorem 3(i) in Li et al. (2016) in the fifth equality, and Theorem 6 in Xiu (2010) along with Assumption A, Theorem 3(i) in Li et al. (2016) in the sixth equality.

\subsection{Proof of Theorem \ref{thEQMLEpowers} (powers of volatility)}
The proof follows the proof of Theorem \ref{thEQMLE} along with the proof of Theorem \ref{thQMLEpowers}.

\subsection{Proof of Theorem \ref{cltuncertainty} (Time-varying friction parameter model with uncertainty zones)}
In order to prove the theorem, we will show that the conditions of Theorem \ref{clt} are satisfied. For this purpose we set $\mathbf{P} = (1,1)$. 

\smallskip
First, Condition (T) follows exactly from Corollary 4.4 (p. 14) in Robert and Rosenbaum (2012). 

\smallskip
We aim to show now Condition (E*). We start with the bias condition (\ref{E611}). To avoid more involved notation, we keep the notation introduced in Section \ref{p2uncertaintyzones} to prove this part. We recall the definition of the estimator of $\eta$ as
\begin{eqnarray*} 
\widehat{\eta}_{t,n} := \sum_{k=1}^{m} \lambda_{t,k,n} u_{t,k,n},
\end{eqnarray*}
with
\begin{eqnarray}
\lambda_{t,k,n} & := & \frac{N_{t,k,n}^{(a)} + N_{t,k,n}^{(c)}}{\sum_{j=1}^{m} \big( 
N_{t,j,n}^{(a)} + N_{\alpha,t,j}^{(c)} \big) },\\
\label{utkn} u_{t,k,n} & := & \max\bigg\{ 0, \min \bigg\{1, \frac{1}{2} \bigg(
k \bigg( \frac{N_{t,k,n}^{(c)}}{N_{t,k,n}^{(a)}} - 1 \bigg) + 1 \bigg) \bigg\} \bigg\}.
\end{eqnarray}
One can see easily from (\ref{utkn}) that $u_{t,k,n}$ are consistent estimators of $\eta$ with bias which satisfies the condition (\ref{E611}). Moreover, as $\widehat{\eta}_{t,n}$ is a linear combination of $u_{t,k,n}$, it also satisfies (\ref{E611}). It remains to show that the estimator of volatility which we recall to be defined as
\begin{eqnarray}
\label{RVtn} \widehat{RV}_{t,n} &= &\sum_{i=1}^{N_n (t)} (\widehat{X}_{\tau_{i,n}} - \widehat{X}_{\tau_{i-1,n}})^2 \text{, where}\\
\label{hatX} \widehat{X}_{\tau_{i,n}} & = & Z_{\tau_{i,n},n} - \alpha_n (1/2 - \widehat{\eta}_{t,n}) \text{sign} (R_{i,n}),
\end{eqnarray}
also satisfies the bias condition. In fact, combining (\ref{RVtn}) and (\ref{hatX}) along with the key relation between $Z_{\tau_{i,n},n}$ and $X_{\tau_{i,n}}$ which can be found in (2.3) on p. 5 in Robert and Rosenbaum (2012), we can deduce that the bias of (\ref{RVtn}) is a function of the bias of $\widehat{\eta}_{t,n}$ which satisfies the condition (\ref{E611}).

\smallskip
We prove now the condition (\ref{E321}). We set an arbitrary $M > 0$. In view of the form of the sampling times (\ref{timesuncertainty}), we have uniformly in $\theta \in K_M$ and in $i=1, \cdots, B_n$ that 
\begin{eqnarray*}
& & \var \big[ h_n^{\frac{1}{2}} \big( \hat{\theta}_{h_n,n} 
  ( R_{i,n}^{1,\mathbf{P},\theta} ; \cdots ; R_{i,n}^{h_n,\mathbf{P}, \theta}) - \theta \big) \Delta \Tau_{i,n}^{\mathbf{P},\theta} \big]\\ 
& = & \var \big[ h_n^{\frac{1}{2}} \big( \hat{\theta}_{h_n,n} 
  (  R_{i,n}^{1,\mathbf{P},\theta} ; \cdots ; R_{i,n}^{h_n,\mathbf{P}, \theta}) - \theta \big) \big] \big( \esp \big[  \Delta \Tau_{i,n}^{\mathbf{P},\theta} \big] )^2\\ & & + o_p (h_n^2 n^{-2}),\\
& = & \var \big[ h_n^{\frac{1}{2}} \big( \hat{\theta}_{h_n,n} 
  (R_{i,n}^{1,\mathbf{P},\theta} ; \cdots ; R_{i,n}^{h_n,\mathbf{P}, \theta}) - \theta \big) \big] \esp \big[  \Delta \Tau_{i,n}^{\mathbf{P},\theta} \big] \Delta \Tau_{i,n}^{\mathbf{P},\theta}  \\ & &+ o_p (h_n^2 n^{-2}),\\
& = & S_{\theta,n}^{(1)}  S_{\theta,n}^{(2)} \Delta \Tau_{i,n}^{\mathbf{P},\theta} T h_n n^{-1} 
 + o_p (h_n^2 n^{-2}),
\end{eqnarray*}
with 
$$S_{\theta,n}^{(1)} := \var \big[ h_n^{\frac{1}{2}} \big( \hat{\theta}_{h_n,n} 
  (R_{i,n}^{1,\mathbf{P},\theta} ; \cdots ; R_{i,n}^{h_n,\mathbf{P}, \theta}) - \theta \big) \big]$$
and $S_{\theta,n}^{(2)} := \esp \big[  \Delta \Tau_{i,n}^{\mathbf{P},\theta} \big] T^{-1} h_n^{-1} n$. By Lemma $4.19$ in p. 26 of Robert and Rosenbaum (2012) in the special case where the volatility is constant, we obtain the existence and the value of $S_{\theta}^{(1)}$ such that $S_{\theta,n}^{(1)} \rightarrow S_{\theta}^{(1)}$. Also, by Corollary $4.4$ in p. 14 of Robert and Rosenbaum (2012), there exists $S_{\theta}^{(2)}$ such that $S_{\theta,n}^{(2)} \rightarrow S_{\theta}^{(2)}$. If we define $V_{\theta} = S_{\theta}^{(1)} S_{\theta}^{(2)}$, (\ref{E321}) is satisfied.

\smallskip
The Lindeberg condition (\ref{E331}) can be obtained using conditional Cauchy-Schwarz inequality, together with the fact that the fourth moment of 
$$h_n^{\frac{1}{2}} \big( \hat{\theta}_{h_n,n} 
  (R_{i,n}^{1,\mathbf{P},\theta} ; \cdots ; R_{i,n}^{h_n,\mathbf{P}, \theta}) - \theta \big)$$
  is bounded, and Condition (T).

\smallskip 
We prove now the conditions (\ref{condnotreg1}) and (\ref{condnotreg2}). Here again we choose the reference martingale $M_t = 0$, and thus we obtain trivially (\ref{condnotreg1}). To show (\ref{condnotreg2}), if we decompose $\big( \hat{\tilde{\Theta}}_{i,n}^{\mathbf{P}} - \theta_{\tau_{i,n}}^* \big)$ following the definition of the estimator, $\Delta \widetilde{\Tau}_{i,n}^{\mathbf{P}}$ as 
$$\sum_{j=1}^{h_n} (\widetilde{\tau}_{(i-1)h_n + j,n}^{\mathbf{P}} - \widetilde{\tau}_{(i-1)h_n + j-1,n}^{\mathbf{P}}),$$
and 
$$N_{\Tau_{i,n}} - N_{\Tau_{i-1,n}} = \sum_{j=1}^{h_n} (N_{(i-1)h_n + j,n}^{\mathbf{P}} - N_{(i-1)h_n + j-1,n}^{\mathbf{P}}),$$
and develop the product of those three expressions, we can easily get rid of the cross terms, and the other terms can be shown going to 0 following the same line of reasoning as the proof of Lemma 4.11 (pp. 20-21) and Lemma 4.14 (pp. 22-23) in Robert and Rosenbaum (2012).

\smallskip
We turn now to (\ref{nonregeq}) and (\ref{nonregeq2}). We start by showing the latter condition. We can decompose $\Delta \widetilde{\Tau}_{i,n}^{\mathbf{P}} - \Delta \Tau_{i,n}$ into
\begin{eqnarray}
\label{condE2tvf} \big( \Delta \widetilde{\Tau}_{i,n}^{\mathbf{P}} - \Delta \breve{\Tau}_{i,n}^{\mathbf{P}} \big) 
+ \big( \Delta \breve{\Tau}_{i,n}^{\mathbf{P}} - \Delta \Tau_{i,n} \big),
\end{eqnarray}
where $\Delta \breve{\Tau}_{i,n}^{\mathbf{P}}$ follows the same definition as $\widetilde{\Tau}_{i,n}^{\mathbf{P}}$ (i.e. we hold volatility constant on the block) except that the starting point of the past is not set to $\mathbf{P}$ but kept to the random past $P_{\Tau_{i-1,n},n}$. We deal with the first term in (\ref{condE2tvf}). We can see that under the parametric model the past $P_{\tau_{i,n}}$ follows a discrete Markov chain on the space $\{1, \cdots, m \} \times \{ -1, 1 \}$. Following the same line of reasoning as in the proof of Lemma $14$ in Potiron and Mykland (2017), we can easily show that 
\begin{eqnarray*}
n^{\frac{1}{2}} \sum_{i=1}^{B_n} \esp_{\Tau_{i-1,n}} \Big[ \big| \Delta \widetilde{\Tau}_{i,n}^{\mathbf{P}} - \Delta \breve{\Tau}_{i,n}^{\mathbf{P}} \big| \Big] & \overset{\proba}{\rightarrow} & 0
  \end{eqnarray*}
We turn now to the second term in (\ref{condE2tvf}). Using the same idea as in the proof of Lemma $11$ in Potiron and Mykland (2017), we deduce
\begin{eqnarray*}
n^{\frac{1}{2}} \sum_{i=1}^{B_n} \esp_{\Tau_{i-1,n}} \Big[ \big| \Delta \breve{\Tau}_{i,n}^{\mathbf{P}} - \Delta \Tau_{i,n} \big| \Big] & \overset{\proba}{\rightarrow} & 0.
  \end{eqnarray*}
We have thus shown that (\ref{nonregeq2}) holds. The same line of reasoning can lead us to (\ref{nonregeq}).

\subsection{Proof of Theorem \ref{exMA1} (Time-varying MA(1))}
The key result for this proof is the connection between the MA(1) process and the observations in the model described in Section \ref{noise} in the case $\alpha=1/2$. Such connection can be seen in view of the proof of Proposition I (pp. 391-393) in A\"{i}t-Sahalia et al. (2005). More specifically, we can use Taylor expansions to re-express this estimator as the estimator in Section \ref{noise} and then use Theorem \ref{thQMLE} (ii) to conclude. Similar Taylor expansions were already obtained in the proof of Theorem \ref{thQMLEpowers}, and we will not further explain the details in this specific case.

\subsection{Estimation of the friction parameter bias and standard deviation in the model with uncertainty zones} \label{apfriction}

\smallskip
In this section, we provide the formal definitions, along with some theoretical derivation, of the friction parameter bias and standard deviation used in our empirical illustration. The notation of Section \ref{p2uncertaintyzones} and Section \ref{empiricalwork} are in force. 

\smallskip 
We estimate the standard deviation as 
$$\widehat{s}_n := \widehat{s}_n (\widehat{\eta}_{T,n}),$$
where a formal expression or an estimator of the variance of $V(\eta) := \widehat{s}_n (\eta)$ is provided in what follows, depending on the setting. We also derive an expression or estimator of the bias of $\widehat{\eta}_{T,n}$, that we call $B(\eta)$. They are both obtained assuming that the friction parameter is fixed to $\eta$. In our numerical study we find that this bias is very close to 0 so that it is relatively safe to assume that it equals 0 for the purpose of statistical inference.

\smallskip
We consider first the case where the absolute jump size is constant equal to the tick size, i.e. $L_{i,n} := 1$, and $N_n(t)$ is non-random. In view of (\ref{hatetamod}), we have $$\widehat{\eta}_{t,n} :=  \min \Big(1, \frac{N_{t,1,n}^{(c)}}{2 N_{t,1,n}^{(a)}} \Big).$$ 
We also have by definition that the number of alternations is $N_{t,1,n}^{(a)} = N_n(t) - N_{t,1,n}^{(c)}$. Then 
\begin{eqnarray} \label{EFP}
N_{t,1,n}^{(c)} \sim Bin (N_n(t), \frac{2 \eta}{2 \eta + 1}), 
\end{eqnarray}
where $Bin(n,p)$ is a binomial distribution with $n$ observations and probability $p$. Let $B \sim Bin (N_n(t), \frac{2 \eta}{2 \eta + 1})$. We can define the bias as
$$B(\eta) := \esp \left[  \min \Big(1, \frac{B}{2 (N_n - B)} \Big) \right] - \eta$$
and the variance as 
$$V(\eta) := \var \left[  \min \Big(1, \frac{B}{2 (N_n - B)} \Big) \right].$$
In this case we have thus shown that $B$ and $V$ can be computed easily numerically. 

\smallskip
We assume now that $N_n(t)$ can be random. We can work conditional on $N_n(t)$. As the sampling times are endogenous, (\ref{EFP}) is not true in that case. Nonetheless, we can still approximate $N_{t,1,n}^{(c)}$ by $Bin (N_n(t), \frac{2 \eta}{2 \eta + 1})$ if the number of observations is large enough.

\smallskip
We now turn out to the general case, i.e. when $L_{i,n}$ can be different from $1$. For $k=1, \cdots, m$ we define $\widetilde{p}_k := \frac{2 \eta + k - 1}{2 \eta + k}$ and we let $B_k$ be an independent sequence of distribution $Bin(N_{t,k,n}^{(c)} + N_{t,k,n}^{(a)}, \widetilde{p}_k)$, and  
$$C_k := \max\Big( 0, \min \Big(1, \frac{1}{2} \big( 
k \big( \frac{B_k}{N_{t,k,n}^{(a)} + N_{t,k,n}^{(c)} - B_k} - 1 \big) + 1 \big) \Big) \Big).$$
The distribution of $\widehat{\eta}_{t,n}$ can be approximated by the distribution of $$\sum_{i=1}^{m} \lambda_{t,k,n} C_k,$$ 
and we can estimate the bias as $\widehat{B}(\eta) := \sum_{i=1}^{m} \lambda_{t,i,n} \esp \big[ C_k \big]$ and the variance as $\widehat{V}(\eta) := \sum_{i=1}^{m} \lambda_{\alpha,t,i}^2 \var \big[ C_k \big]$.

\section{Additional numerical study: the time-varying MA(1) case}
\label{numericalstudyMA1}
\subsection{Goal of the study}
To investigate the finite sample performance of the LPE, we consider the  time-varying MA(1) with null-mean introduced in Section \ref{timeseries}, where the related local estimator is the MLE. The goal of the study is twofold. First, we want to investigate how the LPE performs compared to some naive concurrent approaches. Second, we want to discuss about the choice of the tuning parameter $h_n$ in practice. 

\smallskip
We consider the following simple concurrent approaches:
{\renewcommand\labelitemi{}
\begin{itemize}
  \item \emph{MLE}: the global MLE when considering that the parameters are not time-varying on $[0,T]$.
  \item \emph{Fitting Recent Observations (FRO)}: This approach consists in fitting the MLE on a recent sub-block with less observations (e.g. on $[T_{F}, T]$ where $T_{F} > 0$) so that the parameter is roughly constant on that block. 
\end{itemize}
}
To compute the bias-corrected estimator $\widehat{\Theta}_n^{(BC)} = \widehat{\Theta}_{i,n} - b(\widehat{\Theta}_{i,n}, h_n)$, we can either compute and implement the function $b(\theta,h)$ or carry out Monte-Carlo simulations to compute $b (\theta,h)$ for any $(\theta,h)$ prior to the numerical study. We choose the latter option as this allows to get also rid of bias terms which appears in the Taylor expansion in a higher order than $O(h^{-2})$. Indeed, although those terms vanish asymptotically, they can pop up in a finite sample context. To be more specific, we first compute the sample mean for a grid of parameter values and block length $(\theta,h)$ with, say, 100,000 Monte Carlo paths\footnote{Actually, the number of Monte Carlo paths can be significantly lower when $h$ increases} of the parametric model. Then on each block, we estimate the bias by $b(\widehat{\Theta}_{i,n} , h_n)$.

\smallskip
We discuss here what we expect theoretically from the bias-correction. In view of the decomposition (\ref{decomposeCLT}), we can disentangle the bias of $\widehat{\Theta}_{i,n}$ on first approximation as the sum of two terms, namely the bias of the parametric estimator and the bias due to the fact that the parameter is time-varying. The former can be corrected by the econometrician, and we define the bias-corrected local estimate $\widehat{\Theta}_{i,n}^{(BC)}$ accordingly. On the contrary, as the the parameter path is unknown, we cannot correct for the latter. This is one reason why we have to work with (up to constant terms) a $h_n  < n^{\frac{1}{2}}$. The theory  shows that the normalized latter bias will vanish asymptotically under that condition. Conversely, the econometrician who chooses to work locally with $h_n > n^{\frac{1}{2}}$ will most likely obtain a significant latter bias which she cannot identify, and correcting for the former bias might not improve the estimation in that case.

\subsection{Model design}
\smallskip
We recall that the time-varying parameter is $\theta_t^* = (\beta_t, \kappa_t)$. We set $T=1$, which stands for one day (or one week, one month). We fix the number of observations $n=10,000$. We consider one toy model  where the parameters move around a target parameter deterministically. We assume that the noisy parameter follows a cos function $\theta_t^* = \nu + A \cos (\frac{ 2 \pi t \delta}{T})$, where $\nu = (\beta, \kappa)$ is the parameter, $A=(A^{(\beta)}, A^{(\kappa)})$ corresponds to the amplitude, and $\delta=(\delta^{(\beta)}, \delta^{(\kappa)})$ stands for the number of oscillations on $[0,T]$. With this model, we set $\Theta = (\beta, \kappa)$. We fix the parameter $\nu = (0.5, 1)$ and the amplitude $A=(0.2,0.4)$. We also choose one setting with a small number of oscillations $\delta=(4,4)$ and one with a bigger number of oscillations $\delta=(10,10)$. We simulate $M=1,000$ Monte-carlo repetitions.

\smallskip
In view of Theorem \ref{exMA1} and Condition (P2), the tuning parameter $h_n$ should (up to constant terms) satisfy $n^{1/4} < h_n < n^{1/2}$. In our case, we have that $n^{1/4}=10$ and $n^{1/2}=100$. Accordingly we set $h_n=$ 25, 100, 500, 1000, 2000, 5000. For the FRO approach, we set\footnote{This choice is arbitrary, but different values would yield to similar results.} $T_F = 0.95$, which means that we consider the last 500 observations to fit the MLE.

\subsection{Results}
The results are reported in Table \ref{tableEst1} when $\delta = (4,4)$ and Table \ref{tableEst2} when $\delta = (10,10)$. First, note that the results are similar for both values. Second, as expected from the theory, the LPE performs at its best with the choice $h_n = n^{\frac{1}{2}} = 100$, and the bias-corrected version is much better. Moreover, it outperforms the two concurrent approaches.

\smallskip
The case $h_n = 25$ allows us to check what can happen when we have blocks with very few observations. The bias-corrected estimator performs well to estimate $\kappa$, but somehow the bias-correction to estimate $\beta$ does not provide better estimates. This is most likely due to the fact that we have not enough observations on each block. 

\smallskip
The estimation made with $h_n=500$ is very decent in the case with small number of oscillations. The bias-corrected estimator is actually not as good. This corroborates the theory that when $h_n >> 100$ the main source of bias is due to the parameter which is time-varying rather than the parametric estimator bias itself. If we have a bigger number of oscillations, the estimates are not as accurate. When using bigger $h_n$, we see the same pattern, and the accuracy of the estimation decreases as $h_n$ increases. 

\smallskip
The global MLE performs relatively well to estimate $\beta$, but have a strong bias in $\kappa$. This indicates that even in a simple deterministic model which oscillates around the target value, the MLE cannot be trusted. Finally, the FRO is far off and the standard deviation is bigger. 

\begin{cltblocksize}
 (block size) The conditions of our paper provides the asymptotic order to use for the tuning parameter $h_n$. Thus, it gives a rule of thumb to use on finite sample, but it is 
left to the practitioner to ultimately choose $h_n$. If the parametric estimator is badly biased, the practitioner should increase the value of $h_n$. Also, 
if the parameter seems roughly constant, $h_n$ can be chosen to be bigger. In our simulation study, this rule of thumb could be trusted. In our empirical illustration, we can see that the estimated volatility is robust to the value of $h_n$ if we choose $h_n \approx N_n^{1/2}$. As $n$ can be chosen such that $n =
N_n$, this indicates that the rule of thumb seems to be robust to the actual choice of $h_n$ in our empirical study too.
\end{cltblocksize}

\begin{table}[p]
 \centering
  \begin{tabular}{| l | c | c | c | c | c |}
   \hline
    &  & $\beta$ & $\beta$ & $\kappa$ & $\kappa$ \\
   \hline
    estimator & block size & sample bias & s.d. & sample bias & s.d. \\
  \hline
    MLE & & -0.0052 & 0.0085 & 0.1041 & 0.0148 \\
    FRO & & 0.2913  & 0.0355 & 0.1666 & 0.0355 \\
    LPE & 25 & -0.0168 & 0.0131 & -0.0985 & 0.0096\\
    BC LPE & 25 & -0.0172 & 0.0132 & -0.0062 & 0.0097\\
    LPE & 100 & 0.0035 & 0.0083 & -0.0256 & 0.0096\\
    BC LPE & 100 & -0.0010  & 0.0082 & -0.0065 & 0.0096 \\
    LPE & 500 & -0.0021  & 0.0094 & 0.0073 & 0.0101\\
    BC LPE & 500 & -0.0049 & 0.0095 & 0.0098 & 0.0104\\
    LPE & 1000 & -0.0030  & 0.0099 & 0.0425 & 0.0125\\
    BC LPE & 1000 & -0.0056 & 0.0100 & 0.0438 & 0.0126\\
    LPE & 2000 & -0.0032  & 0.0102 & 0.1029 & 0.0143\\
    BC LPE & 2000 & -0.0055  & 0.0101 & 0.1035 & 0.0143\\
    LPE & 5000 & -0.0052  & 0.0087 & 0.1037 & 0.0148\\
    BC LPE & 5000 & -0.0060  & 0.0087 & 0.1044 & 0.0147 \\
    
   \hline
  \end{tabular}
  \caption{In this table, we report the sample bias and the standard deviation for the different estimators in the case of a small number of oscillations $\delta=(4,4)$. The parameter $(\beta, \kappa) = (0.5,1)$. The number of Monte-carlo simulations is 1,000. Note that BC stands for "bias-corrected".}
\label{tableEst1}
 \end{table}
 
 \begin{table}[p]
 \centering
  \begin{tabular}{| l | c | c | c | c | c |}
   \hline
    &  & $\beta$ & $\beta$ & $\kappa$ & $\kappa$ \\
   \hline
    estimator & block size & sample bias & s.d. & sample bias & s.d. \\
  \hline
    MLE & & -0.0069 & 0.0105 & 0.1094 & 0.0222 \\
    FRO & & 0.0065  & 0.0391 & 0.0882 & 0.0678 \\
    LPE & 25 & -0.0148 & 0.0183 & -0.0876 & 0.0144\\
    BC LPE & 25 & -0.0155 & 0.0184 & 0.0046 & 0.0143\\
    LPE & 100 & 0.0017 & 0.0092 & -0.0164 & 0.0183\\
    BC LPE  & 100 & 0.0012  & 0.0092 & 0.0039 & 0.0183\\
    LPE & 500 & -0.0053  & 0.0094 & 0.1046 & 0.0219\\
    BC LPE  & 500 & -0.0085 & 0.0094 & 0.1086 & 0.0221\\
    LPE & 1000 & -0.0071  & 0.0102 & 0.1078 & 0.0216\\
    BC LPE  & 1000 & -0.0115 & 0.0102 & 0.1098 & 0.0217\\
    LPE & 2000 & -0.0071  & 0.0106 & 0.1087 & 0.0220\\
    BC LPE & 2000 & -0.0108  & 0.0106 & 0.1096 & 0.0221\\
    LPE & 5000 & -0.0071  & 0.0106 & 0.1087 & 0.0220\\
    BC LPE & 5000 & -0.0093  & 0.0106 & 0.1090 & 0.0219 \\
    
   \hline
  \end{tabular}
  \caption{In this table, we report the sample bias and the standard deviation for the different estimators in the case of a bigger number of oscillations $\delta=(10,10)$. The parameter $(\beta, \kappa) = (0.5,1)$. The number of Monte-carlo simulations is 1,000. Note that BC stands for "bias-corrected".}
\label{tableEst2}
 \end{table}
 
 \section{Empirical illustration in the model with uncertainty zones} \label{empiricalwork}
In this section, we implement the LPE in the model with uncertainty zones introduced in Section \ref{p2uncertaintyzones}. We recall that the parameter of interest is defined as $\xi_t^* = (\sigma_t^2, \eta_t)$. We are looking at Orange (ORA.PA) stock price traded actively on the CAC 40 on one random day, Monday March 4th, 2013. To prevent from opening and closing effect, we assume that restrict to data obtained from 9am to 4pm. The number of transactions inducing to a price change during this time period is equal to $N_n = 3306$, the tick size $\alpha_n = 0.001$ euro, and the price is equal to $8$ euros on average.

\begin{figure}
\includegraphics[width=\linewidth]{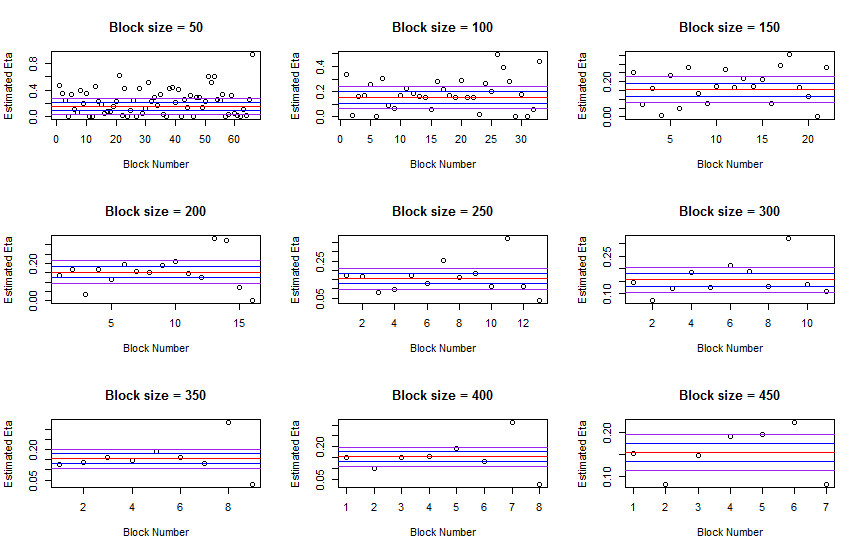}
\centering
\caption{Estimated friction parameter over time for different values of $h_n$. The red line corresponds to the global estimate. The blue lines are one (local) standard deviation away from the global estimator. The purple lines are two (local) standard deviations away.}
\label{emp1}
\end{figure}

\smallskip
We report the global estimate $\widehat{\eta}_{T,n} = 0.155$, and the standard deviation\footnote{related definitions and derivation of $\widehat{s}_n$ and $\hat{s}_{i,n}$ can be found in Section \ref{apfriction}} $\widehat{s}_n = 0.008$. Moreover, Figure \ref{emp1} documents the friction parameter estimates over time for several values of block size. Based on those estimates and the local standard deviation estimate $\hat{s}_{i,n}$, we compute the associated chi-square statistic\footnote{Note that since the number of observations of the last block is arbitrary, the last block estimate is not used to compute the chi-square statistic.} $$\chi_n^2 := \sum_{i=1}^{B_n - 1} \Big( \frac{\hat{\eta}_{i,n} - \widehat{\eta}_{T,n}}{\hat{s}_{i,n}} \Big)^2.$$
Under the null hypothesis which states that $\eta_t$ is constant, $\chi_n^2$ follows approximately a chi-square distribution with $B_{n}-1$ degrees of freedom. We report $\chi_n^2$ for different values of $h_n$ in Table \ref{chisq}. The obtained values indicate that we have strong evidence against the null hypothesis, revealing that the friction parameter is time-varying.\footnote{This analysis has been carried out on other days and other stocks. We consistently conclude that the friction parameter is time-varying.}

\begin{figure}
\includegraphics[width=.5\linewidth]{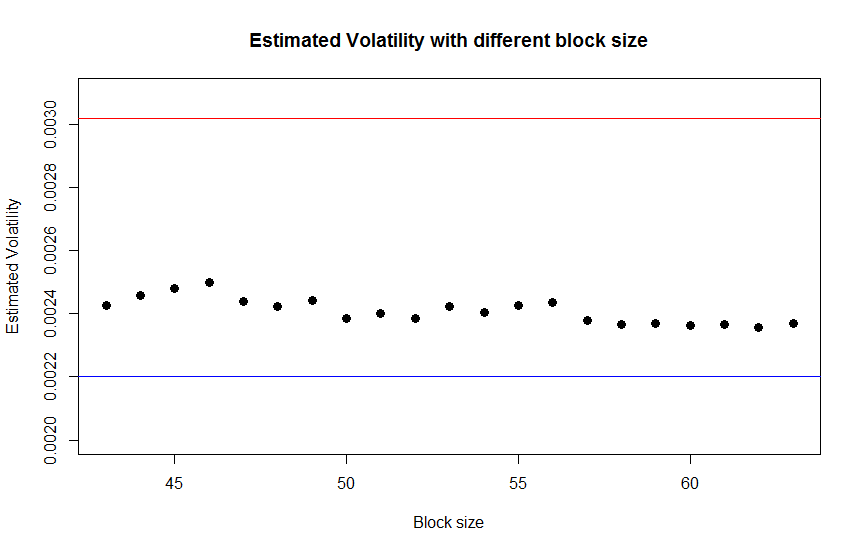}
\centering
\caption{Estimated volatility with the LPE for different values of $h_n$. The red line corresponds to the RV estimator. The blue line stands for the global model with uncertainty zones volatility estimator.}
\label{emp2}
\end{figure}

\smallskip 
We report in Figure \ref{emp2} the estimated volatility for different values of $h_n$. Because $N_n^{1/2} \approx 57.5$ we set $h_n =  43, \cdots , 63$. We also report the estimates with RV and the global model with uncertainty zones. The estimates of the latter seems to slightly underestimate the integrated volatility. In addition, the former, which is positively biased under the presence of microstructure noise, is far off, most likely overestimating the integrated volatility. Finally, the estimates are very similar for different values of $h_n$, which seems to indicate that the method is robust to small block size variation.

\begin{table}
 \centering
  \begin{tabular}{ | c | c | c | c | c |}
   \hline
    $h_n$ & $B_n$  & Chi Sq. Stat & Dg. Fr. & p-value\\
  \hline
   $50$ & $67$ & $719$ & $66$ & $\approx 0$\\
   $100$ & $34$ & $268$ & $33$ & $\approx 0$\\
   $150$ & $23$ & $155$ & $22$ & $\approx 0$\\
   $200$ & $17$ & $116$ & $16$ & $\approx 0$\\
   $250$ & $14$ & $109$ & $13$ & $\approx 0$\\
   $300$ & $12$ & $68.5$ & $11$ & $\approx 0$\\
   $350$ & $10$ & $90.6$ & $9$ & $\approx 0$\\
   $400$ & $9$ & $91.5$ & $8$ & $\approx 0$\\
   $450$ & $8$ & $42.6$ & $7$ & $\approx 0$\\
   \hline
  \end{tabular}
  \caption{Summary chi-square statistics $\chi_n^2$ based on the block size $h_n$.} 
\label{chisq}
 \end{table}

\end{document}